\documentclass[aps,onecolumn,showpacs]{revtex4}
\usepackage{dcolumn}
\usepackage{graphicx}
\usepackage{amsmath}
\usepackage{amsfonts}
\usepackage{amssymb}
\usepackage{psfrag}
\usepackage{wrapfig}
\usepackage{subfigure}
\usepackage{makeidx}
\usepackage{bm}
\usepackage{epsf}

\newtheorem{thm}{Theorem}

 \newtheorem{lem}{Lemma}
 \newtheorem{prop}{Proposition}

\begin{document}

\title{Darboux transformation and multi-dark soliton for N-component coupled nonlinear Schr\"odinger equations}
\author{Liming Ling$^1$}
\author{Li-Chen Zhao$^2$}
\author{Boling Guo$^3$}
\address{$^1$School of Sciences, South China University of Technology, Guangzhou 510640, China}
\address{$^2$Department of Physics, Northwest University, 710069, Xi'an, China}
\address{$^3$Institute of Applied Physics and Computational Mathematics, 100088, Beijing, China}

\begin{abstract}
In this paper, we obtain a uniform Darboux transformation for multi-component coupled NLS equations, which can be reduced to all previous presented Darboux transformation.
As a direct application, we derive the single dark soliton and multi-dark soliton solutions for multi-component coupled  NLS with defocusing case and mixed focusing and defocusing case.
Some exact single and two-dark solitons of three-component NLS equation are shown by plotting the picture.

\


\
\end{abstract}

 \maketitle
\section{Introduction}
It is well known that nonlinear Schr\"odinger (NLS)-type equations play a prominent role in nonlinear physical systems, such as nonlinear optics \cite{Agra},
Bose-Einstein condensates \cite{DGPS}. In these physical systems,
 the nonlinear coefficient can be positive or negative, depending on the
physical situations \cite{Ablowitz}. For example, the nonlinearity can be positive or negative when  the interaction between atoms is repulsive or attractive in Bose-Einstein condensates
\cite{DGPS}. For nonlinear optics, it corresponds to the focusing or defocusing case. There are lots of results about the scalar NLS equation \cite{ZS}. It admits bright solitons \cite{ZS}, breather \cite{MA} and rogue wave \cite{Pere} in the focusing
case, and dark solitons \cite{FT} in the defocusing case.

Since a variety of complex systems, such as Bose-Einstein
condensates, nonlinear optical fibers, etc., usually involve
more than one component, the studies were extended to coupled NLS equations cases. The integrable coupled NLS equations are called Manakov system \cite{Manakov}. In the focusing case,
the coupled NLS equations admit
bright-bright solitons, bright-dark, breather, rogue wave, bright-dark-breather and bright-dark-rogue wave solution \cite{WZY,BDCW,GL1}. In the defocusing case, the coupled NLS equations admit bright-dark and dark-dark
solitons \cite{SK,RL}. In the mixed case,
the coupled NLS equations
admit bright-bright solitons, bright-dark solitons and dark-dark soliton \cite{OWY,VKL}. For the focusing three-component nonlinear Schr\"odinger equations,
there exists the ``four-petaled flower" structure rogue wave by Darboux transformation (DT) \cite{ZL}.

The defocusing NLS equation possesses dark soliton. However, the dark soliton is not evident to research by Darboux transformation method.
The first time to obtain the dark soliton with DT method was appeared in 1996 \cite{Manas}. The single dark soliton formula was appeared in 2006 and 2009
\cite{DL,DL1}. The method for inverse scattering method is even more open until 2006 \cite{ABP}. Even the method in reference \cite{ABP}, their solutions can be degenerated into scalar NLS equation.

In this paper, we will derive a simple N-soliton formula for dark soliton and breather solution of integrable multi-component coupled  system. In order to present our work clearly and readably, we revisit the method in 1996 and 2006 \cite{Manas,DL}. In 1996 \cite{Manas}, Manas proposed a method to derive the dark soliton for dNLS. The Darboux matrix can be represented as
\begin{equation*}
    T=I+\frac{\mu_1-\lambda_1}{\lambda-\lambda_1}\frac{\Phi_1\Psi_1}{\Psi_1\Phi_1}
\end{equation*}
where $\mu_1$ and $\lambda_1$ are real numbers, $\Phi_1$ is a solution with spectral parameter $\lambda=\lambda_1$, $\Psi_1$ is a solution of conjugation system with spectral parameter $\lambda=\mu_1.$ One fold DT could yields two dark solion for dNLS. The symmetry relation is giving as
\begin{equation*}
    T^{\dag}(\lambda^*)\Sigma T(\lambda)=f(\lambda)\Sigma.
\end{equation*}
However, try our best, the method can not be applied to multi-component case beside the degenerate case.

In \cite{DL}, Degasperis and Lambardo presented a one fold DT method for dark soliton. The Darboux matrix
\begin{equation}\label{DL}
    D(x,t;\lambda)=I+{\rm i}\frac{\hat{z}\hat{z}^{\dag}\Sigma}{(\lambda-\lambda_1)P_1},\quad \hat{z}^{\dag}\Sigma\hat{z}=0,\quad P_1=\int \hat{z}^{\dag}\Sigma\hat{z}dx.
\end{equation}
Although we can iterate the DT method step by step. However, the integration restrict us to do this.

In 2009 \cite{C}, Cie\'sli\'nski revisit different type DT method. Indeed, we can see that the DT \eqref{DL} is nothing but nilpotent Darboux matrix.
However, even though we find this, we can not still iterate the DT with a convenient way. In order to provide a simple way to derive the N-fold DT, we
need some new idea. Firstly, we convert equation \eqref{DL} into the classical binary DT. The important property for DT is the kernel for $D(x,t;\lambda)$:
\begin{equation*}
    \lim_{\lambda\rightarrow \lambda_1}D(x,t;\lambda_1)\hat{z}(\lambda)=0.
\end{equation*}
Indeed, we have the following equality
\begin{equation*}
    \lim_{\lambda\rightarrow \lambda_1}\frac{{\rm i}\hat{z}_1^{\dag}\Sigma\hat{z}(\lambda)}{\lambda-\lambda_1}=-\int \hat{z}_1^{\dag}\Sigma\hat{z}_1dx.
\end{equation*}
Thus the Darboux transforation can be represented as
\begin{equation*}
    \hat{z}(\lambda)\rightarrow D(x,t;\lambda)\hat{z}(\lambda)=\hat{z}(\lambda)-\frac{\hat{z}_1\Omega(\hat{z}_1,\hat{z})}{\Omega(\hat{z}_1,\hat{z}_1)},\quad d\Omega(f,g)=f^{\dag}\Sigma g dx.
\end{equation*}
We can see that this transformation is nothing but classical binary Darboux transforation. However, generally speaking, Darboux transformation is considered as the special gauge transformation. That is the reason that we underestimate the classical binary DT. Indeed, the binary DT is the consistent transformation for NLS type system. With the binary DT method. We can reduce it for Zakharov-Shabat dressing operator or the loop group representation.
What is important, we can obtain the dark soliton's DT.

Recently, there are some another method to derive the dark soliton for multi-component NLS equations. For instance, the algebraic-geometry reduction method
 and dressing-Hirota method \cite{ABS,Kalla}.

In this paper, we depart from the classical binary Darboux transformation of AKNS system. To the best of our knowledge, although the binary DT for KdV, KP and DS equations are well known to us, there no research about the binary
DT for AKNS system. Since in classical view, they deem that the DT of AKNS system is the gauge transformation \cite{Gu}, loop group \cite{TU} or
Zaharov-Shabat dressing operator \cite{NMPZ}. All of above transformation can be diagonalize, but the binary DT can not be diagonalize, since the
the operator is nilpotent. In order to solve this problem in frame of diagonalize, we use the limit technique.

The aim of our work is two fold. Firstly, we reduce the binary DT of AKNS system to obtain an uniform transformation for AKNS system.
Secondly, we use the binary DT of AKNS system to derive dark soliton and multi-dark soliton for multi-component NLS equations. In section \ref{sec2},
we introduce some basic knowledge about AKNS system. Then we emphasize the binary DT for AKNS system. Based on the binary DT, we reduce this transformation with an uniform transformation for AKNS system. In section \ref{sec3}, by the uniform transformation, we derive the single dark soliton and multi-dark soliton for N-component NLS equation. In order to give us a clear understand of our formula, we plot the explicit dark soliton and two dark soliton picture of three-component NLS euqations. Final section involves some conclusions and discussions.

\section{The AKNS system and the Binary Darboux transformation}\label{sec2}

In this section, we firstly recall some results about the AKNS system and its reduction for NLS equations. Secondly, we introduce the
binary DT for AKNS system. Finally, we reduce the binary DT into different transformation by conjugation equation and limit technique. Through the limit
technique, the integration is automatic by taking the limit.

\subsection{The AKNS system}
In this subsection, we recall the classical results about the AKNS system \cite{AKNS}. Let $a=\mathrm{diag}(a_1,a_2,\cdots,a_n)$ be a fixed nonzero diagonal matrix in $\mathrm{sl}(n,\mathbb{C}),$ and denote
\begin{equation*}
    \begin{split}
      {\rm sl}(n)_a= &\{y\in {\rm sl}(n,\mathbb{C})|[a,y]=0\}, \\
        {\rm sl}(n)_a^{\bot}=&\{y\in {\rm sl}(n,\mathbb{C})|{\rm tr}(zy)=0 \quad \text{for any }z\in {\rm sl}(n)_a  \}.
    \end{split}
\end{equation*}
Let $L^{\infty}(\mathbb{R},{\rm sl}(n)_a^{\bot})$ denotes the space of maps in the $L^{\infty}(\mathbb{R})$ class. For the spectral problem
\begin{equation}\label{Akns}
    \Phi_x=(a\lambda+u(x))\Phi,
\end{equation}
when $\lambda\rightarrow \infty$, we have the formal asymptotical behavior
\begin{equation*}
    \Phi\rightarrow \exp[a\lambda x].
\end{equation*}
Thus we can suppose $\Phi=m(x;\lambda){\rm e}^{a\lambda x}$, where
$m(x;\lambda)$ is an analytical function and possess the following
formal expansion
\begin{equation*}
m(x;\lambda)=I+m_1(x)\lambda^{-1}+m_2(x)\lambda^{-2}+\cdots.
\end{equation*}
 Substituting it into \eqref{Akns}, we can obtain
\begin{equation*}
    \Phi_x\Phi^{-1}=mam^{-1}\lambda+m_xm^{-1}=a\lambda+[m_1,a]+O(\lambda^{-1})
\end{equation*}
and $\Phi_x\Phi^{-1}$ being holomorphic in $\lambda\in \mathbb{C}$
imply that $\Phi_x\Phi^{-1}=a\lambda +[m_1,a].$

Let $b\in{\rm sl}(n,\mathbb{C})$ such that $[b,a]=0$, we have the allowing formal expansion of $mbm^{-1}$ at $\lambda=\infty$,
\begin{equation*}
    mbm^{-1}\sim Q_{b,0}+Q_{b,1}\lambda^{-1}+Q_{b,2}\lambda^{-2}+\cdots.
\end{equation*}
Since $\Phi_x\Phi^{-1}=\lambda a +u$ and $\Phi b\Phi^{-1}=mbm^{-1}$,
we can obtain that
\begin{equation*}
    \left[\partial_x+a\lambda+u,\Phi b\Phi^{-1}\right]=0.
\end{equation*}
It follows that
\begin{equation}\label{flow}
(Q_{b,j}(u))_x+[u,Q_{b,j}(u)]=[Q_{b,j+1}(u),a].
\end{equation}
Write
\begin{equation*}
    Q_{b,j}=T_{b,j}+P_{b,j}\in {sl}(n)_a+{\rm sl}(n)_a^{\bot}.
\end{equation*}
Then equation \eqref{flow} gives
\begin{equation}\label{flow1}
    \begin{split}
      P_{b,j}&=-{\rm ad}(a)^{-1}((P_{b,j-1})_x+\pi_1([u,Q_{b,j-1}])), \\
       (T_{b,j})_x &=-\pi_0([u,P_{b,j}]),
    \end{split}
\end{equation}
In reference \cite{TU}, they proved that if $b$ is a polynomial of
$a$, then $Q_{b,j}$ is an order-(j-1) polynomial differential
operator in $u.$

Then we have the following proposition:
\begin{prop}
Suppose $u(\cdot,t)\in L^{\infty}(\mathbb{R},{\rm sl}(n)_a^{\bot})$
for all $t$,
\begin{equation*}
    \left[\partial_x+a\lambda+u,\partial_t+b\lambda^j+v_1\lambda^{j-1}
    +\cdots+v_j\right]=0
\end{equation*}
for some $v_1,\cdots,v_j,$ and $\lim_{x\rightarrow -\infty}v_{k}(x,t)=\lim_{x\rightarrow-\infty}Q_{b,k}(u(x,t))$
for all $1\leq k\leq j.$ Then we have $v_k=Q_{b,k}(u),$
and
\begin{equation}\label{flow2}
    u_t=(Q_{b,j}(u))_x+[u,Q_{b,j}(u)]=[Q_{b,j+1}(u),a].
\end{equation}
\end{prop}

In what following, we consider the reality conditions. The detail
for reality conditions are given in reference \cite{TU}. A Lax pair
$[\partial_x+A(x,t;\lambda),\partial_t+B(x,t;\lambda)]=0$ is said to
satisfy the reality condition if
$\sigma(A(x,t;\bar{\lambda}))=A(x,t;\lambda)$ and
$\sigma(B(x,t;\bar{\lambda}))=B(x,t;\lambda),$ where $\sigma$ is
complex conjugate linear Lie algebra involution in ${\rm
sl}(n,\mathbb{C})$.

In this paper, we merely consider the $u(k,N+1-k)$ hierarchy \cite{TU}. The second flow in $u(k,N+1-k)$ hierarchy is the following
multi-component NLS equation.
\begin{equation}\label{dcnls}
    {\rm i}\mathbf{q}_t+\frac{1}{2}\mathbf{q}_{xx}-\mathbf{q}\mathbf{q}^{\dag}S\mathbf{q}=0,
\end{equation}
where
\begin{equation*}
    \begin{split}
    \mathbf{q}&=(q_1,\,q_2,\cdots,q_N)^T,\\
    S&=\mathrm{diag}\left(s_1,\,s_2,\cdots,\,s_N\right),\quad s_i=1,
   \text{ for } 1\leq i\leq k,s_i=-1, \text{ for } k+1\leq i\leq N,
    \end{split}
\end{equation*}
 which admits the following Lax pair
\begin{equation}\label{L-P}
    \begin{split}
      \Phi_x &=\left({\rm i}\lambda \sigma_3+{\rm i}Q\right)\Phi,  \\
       \Phi_t &=\left({\rm i}\lambda^2\sigma_3+{\rm i}\lambda Q-\frac{1}{2}({\rm i}\sigma_3Q^2-\sigma_3Q_x)\right)\Phi,
    \end{split}
\end{equation}
where
\begin{equation*}
    Q=\begin{bmatrix}
        0 & -\mathbf{q}^{\dag}S \\
        \mathbf{q} & \mathbf{0}_{N\times N} \\
      \end{bmatrix},\quad \sigma_3=\begin{bmatrix}
                                     1 & \mathbf{0}_{1\times N} \\
                                     \mathbf{0}_{N\times 1} & -\mathbf{I}_{N\times N} \\
                                   \end{bmatrix}.
\end{equation*}
\subsection{The Binary Darboux transformation for AKNS system}
We consider the binary Darboux transformation for AKNS system with symmetry reduction.
Firstly, we give some lemmas.
\begin{lem}\label{lem1}
Suppose $\Phi_1$ and $\Phi$ are the special vector solutions for system \eqref{L-P} at $\lambda=\lambda_1$ and $\lambda$ respectively, then we can have total differention
\begin{equation}\label{omega}
    d\Omega(\Phi_1,\Phi)=\Phi_1^{\dag}\Lambda\sigma_3\Phi dx+\left[(\lambda+\bar{\lambda_1})\Phi_1^{\dag}\Lambda\sigma_3\Phi+\Phi_1^{\dag}\Lambda Q\Phi\right]dt.
\end{equation}
In addition, we have
\begin{equation}\label{Omega1}
    \Omega(\Phi_1,\Phi)= \frac{\Phi_1^{\dag}\Lambda\Phi}{{\rm i}(\lambda-\bar{\lambda_1})}+C.
\end{equation}
If $\lambda_1\in \mathbb{R}$, we have
\begin{equation}\label{Omega2}
    \Omega(\Phi_1,\Phi_1)=\lim_{\lambda\rightarrow \lambda_1}\frac{\Phi_1^{\dag}\Lambda\Phi}{{\rm i}(\lambda-\lambda_1)}+C,
\end{equation}
where $C$ is a complex constant, and $\Lambda=\mathrm{diag}\left(1,\Lambda_1,\,\cdots,\Lambda_k,\,1,\cdots,1\right)$,
$\Lambda_i=-1$, $1\leq i\leq k$.
\end{lem}
\textbf{Proof:}
Taking complex conjugation to \eqref{L-P} both sides, we have
\begin{equation}\label{L-P1}
    \begin{split}
      \Phi_{1,x}^{\dag}\Lambda &=-\Phi_1^{\dag}\Lambda[{\rm i}\bar{\lambda_1}\sigma_3+{\rm i}Q],  \\
      \Phi_{1,t}^{\dag}\Lambda &=-\Phi_1^{\dag}\Lambda[{\rm i}\bar{\lambda_1}^2\sigma_3+{\rm i}\bar{\lambda_1} Q-\frac{1}{2}\sigma_3({\rm i}Q^2-Q_x)].
    \end{split}
\end{equation}
Left multiplying by $\Phi_1^{\dag}\sigma_3$ into both sides of \eqref{L-P} and right multiplying by $\Phi$ into both sides of \eqref{L-P1}, then can can obtain
\begin{equation*}\begin{split}
            \left[\frac{\Phi_1^{\dag}\Lambda\Phi}{{\rm i}(\lambda-\bar{\lambda_1})}\right]_x&=\Phi_1^{\dag}\Lambda\sigma_3\Phi,  \\
             \left[\frac{\Phi_1^{\dag}\Lambda\Phi}{{\rm i}(\lambda-\bar{\lambda_1})}\right]_t&=(\lambda+\bar{\lambda_1})\Phi_1^{\dag}\Lambda\sigma_3
             \Phi+\Phi_1^{\dag}\Lambda Q\Phi.
                 \end{split}
\end{equation*}
It follows that equations \eqref{omega} and \eqref{Omega1} are verified.
$\square$

In what following, to keep the uniqueness of the constants $C$, we
choose it as zero. Following the idea in the introduction, we can
obtain that the one fold binary Darboux transformation for dCNLS
\eqref{dcnls} is
\begin{equation}\label{DT}
\begin{split}
\Phi(\lambda)&\rightarrow \Phi[1]=\Phi-\frac{\Phi_1\Omega(\Phi_1,\Phi)}{\Omega(\Phi_1,\Phi_1)}, \\
Q&\rightarrow Q[1]=Q-{\rm i}\left[\sigma_3,\frac{\Phi_1\Phi_1^{\dag}\Lambda}{\Omega(\Phi_1,\Phi_1)}\right].
\end{split}
\end{equation}
In the following, we verify the validity of the above
transformation.
\begin{thm}
Suppose $\Phi$ satisfies the system \eqref{L-P}, and $\Phi_1$ is a special solution for Lax pair \eqref{L-P} at $\lambda=\lambda_1$,
and $\Phi_1^{\dag}\Lambda\Phi_1=0$ if $\lambda_1\in\mathbb{R}$, then we have
\begin{equation}\label{L-P-N}
    \begin{split}
     \Phi[1]_x &=\left({\rm i}\lambda \sigma_3+{\rm i}Q[1]\right)\Phi[1],  \\
       \Phi[1]_t &=\left({\rm i}\lambda^2\sigma_3+{\rm i}\lambda Q[1]-\frac{1}{2}({\rm i}\sigma_3Q[1]^2-\sigma_3Q[1]_x)\right)\Phi[1].
    \end{split}
\end{equation}
\end{thm}
\textbf{Proof:}
We firstly verify the first equation of \eqref{L-P-N}. By lemma \ref{lem1}, we have $\Omega(\Phi_1,\Phi)=\frac{\Phi_1^{\dag}\Lambda\Phi}{{\rm i}(\lambda-\bar{\lambda_1})}.$ It follows that
\begin{equation*}
    \begin{split}
      \Phi[1]_x=&({\rm i}\lambda \sigma_3+{\rm i}Q)\Phi-\frac{({\rm i}\lambda_1 \sigma_3
      +{\rm i}Q)\Phi_1\Omega(\Phi_1,\Phi)}{\Omega(\Phi_1,\Phi_1)}-\frac{\Phi_1\Phi_1^{\dag}\Lambda\sigma_3\Phi}{\Omega(\Phi_1,\Phi_1)}
      +\frac{\Phi_1\Phi_1^{\dag}\Lambda\sigma_3\Phi_1\Omega(\Phi_1,\Phi)}{\Omega^2(\Phi_1,\Phi_1)}  \\
       =&\left[{\rm i}\lambda\sigma_3+{\rm i}\left(Q-{\rm i}\frac{\sigma_3\Phi_1\Phi_1^{\dag}\Lambda}{\Omega(\Phi_1,\Phi_1)}+{\rm i}\frac{\Phi_1\Phi_1^{\dag}\Lambda\sigma_3}{\Omega(\Phi_1,\Phi_1)}\right)\right]\left(\Phi-\frac{\Phi_1\Omega(\Phi_1,\Phi)}{\Omega(\Phi_1,\Phi_1)}\right),
    \end{split}
\end{equation*}
where the second equality use the relation $\Phi_1^{\dag}\Lambda\Phi_1=0$ if $\lambda_1\in\mathbb{R}$. And it is readily to verify the validity of symmetry relation for $Q[1].$
Then the first equation of \eqref{L-P-N} is verified. Besides, we can obtain the following relation. Since $\Phi[1]=T\Phi,$ $$T=I-\frac{\Phi_1\Phi_1^{\dag}\Lambda}{{\rm i}(\lambda-\bar{\lambda_1})\Omega(\Phi_1,\Phi_1)},$$ it follows that
 \begin{equation}\label{Ttran}
    T_x+{\rm i}T(\lambda\sigma_3+Q)={\rm i}(\lambda\sigma_3+Q[1])T.
\end{equation}
Expanding the matrix $T$ with $T=I+\frac{T_1}{\lambda}+\frac{T_2}{\lambda^2}+\cdots$ and substituting into equation \eqref{Ttran}. Comparing the coefficient of $\lambda$, we can get
\begin{equation}\label{coeff1}
    Q[1]T_1=T_1Q+[T_2,\sigma_3]-{\rm i}T_{1,x}.
\end{equation}

Finally, we consider the time evolution equation of \eqref{L-P-N}. Since matrix $T$ is a special gauge transformation, by directly calculating we have
 \begin{equation}\label{Ttran-t}
    \left[T_t+T\left({\rm i}\lambda^2\sigma_3+{\rm i}\lambda Q-\frac{1}{2}\sigma_3({\rm i}Q^2-Q_x)\right)\right]T^{-1}={\rm i}\lambda^2\sigma_3+{\rm i}\lambda Q[1]+V_1[1],
\end{equation}
where $V_1[1]=-\frac{1}{2}({\rm i}\sigma_3Q^2-\sigma_3Q_x)+{\rm i}[T_2,\sigma_3]+{\rm i}(T_1Q-Q[1]T_1).$ By equation \eqref{coeff1}, we have $V_1[1]^{o}=\frac{1}{2}\sigma_3Q[1]_x,$ and then we have $V_1[1]^{d}=-\frac{1}{2}\sigma_3Q[1]^2$ by directly calculating,
where the superscript $^{o}$ and $^{d}$ denotes the off-diagonal and diagonal part of the block matrix respectively. This completes the theorem.
$\square$

In the following, we consider the N-fold binary Darboux transformation based on above theorem.

\begin{thm}
Suppose we have $N$ different solutions $\Phi_i$s  for Lax pair \eqref{L-P} at $\lambda=\lambda_i$ $(i=1,2,\cdots,N)$,  and $\Phi_j^{\dag}\sigma_3\Phi_j=0$ if $\lambda_j\in \mathbb{R}$, then
the N-fold binary Darboux transformation can be represented as
\begin{equation}\label{N-fold}
    \Phi\rightarrow \Phi[N]=\Psi-NM^{-1}\Omega,\quad N=\left(\Phi_1,\Phi_2,\cdots,\Phi_N\right),
\end{equation}
where
\begin{equation*}
    M=\begin{bmatrix}
        \Omega(\Phi_1,\Phi_1) & \Omega(\Phi_1,\Phi_2) & \cdots & \Omega(\Phi_1,\Phi_N) \\
        \Omega(\Phi_2,\Phi_1) & \Omega(\Phi_2,\Phi_2) & \cdots & \Omega(\Phi_2,\Phi_N) \\
        \vdots & \vdots & \ddots & \vdots \\
        \Omega(\Phi_N,\Phi_1) & \Omega(\Phi_N,\Phi_2) & \cdots & \Omega(\Phi_N,\Phi_N) \\
      \end{bmatrix},\quad \Omega=\begin{bmatrix}
                                   \Omega(\Phi_1,\Phi) \\
                                   \Omega(\Phi_2,\Phi) \\
                                   \vdots \\
                                   \Omega(\Phi_N,\Phi) \\
                                 \end{bmatrix}.
\end{equation*}
The transformation between fields is
\begin{equation*}
    Q\rightarrow Q[N]=Q-{\rm i}\left[\sigma_3,NM^{-1}N^{\dag}\Lambda\right].
\end{equation*}
\end{thm}
\textbf{Proof:}
In the first place, we have
\begin{equation*}
    \begin{split}
       \Phi[N]_x&=\Phi_x-N_xM^{-1}\Omega-NM^{-1}N_x+NM^{-1}M_xM^{-1}N  \\
       &=({\rm i}\lambda\sigma_3+{\rm i}Q)\Phi-{\rm i}(QN+\sigma_3ND)M^{-1}\Omega-NM^{-1}N^{\dag}\Phi+NM^{-1}N^{\dag}NM^{-1}\Omega\\
       &=({\rm i}\lambda\sigma_3+{\rm i}Q)\Phi-{\rm i}QNM^{-1}\Omega-NM^{-1}N^{\dag}\Phi+NM^{-1}N^{\dag}NM^{-1}\Omega\\
       &\quad +{\rm i}\sigma_3NM^{-1}(-\lambda+D)\Omega+\sigma_3NM^{-1}N^{\dag}\sigma_3\Phi-{\rm i}\sigma_3NDM^{-1}\Omega\\
       &=({\rm i}\lambda\sigma_3+{\rm i}Q)\Phi-{\rm i}QNM^{-1}\Omega-NM^{-1}N^{\dag}\Phi+NM^{-1}N^{\dag}NM^{-1}\Omega\\
       &\quad -{\rm i}\lambda\sigma_3NM^{-1}+\sigma_3NM^{-1}N^{\dag}\sigma_3\Phi-{\rm i}\sigma_3NM^{-1}({\rm i}MD-{\rm i}DM)M^{-1}\Omega\\
       &=({\rm i}\lambda\sigma_3+{\rm i}Q)\Phi-{\rm i}QNM^{-1}\Omega-NM^{-1}N^{\dag}\Phi+NM^{-1}N^{\dag}NM^{-1}\Omega\\
       &\quad -{\rm i}\lambda\sigma_3NM^{-1}+\sigma_3NM^{-1}N^{\dag}\sigma_3\Phi-{\rm i}\sigma_3NM^{-1}N^{\dag}\sigma_3 NM^{-1}\Omega\\
       &=({\rm i}\lambda\sigma_3+{\rm i}Q[N])\Phi[N],
    \end{split}
\end{equation*}
where the third equality uses the relation $(-\lambda+D)\Omega+N^{\dag}\sigma_3\Phi=0$, the fifth equality uses the relation $N^{\dag}\sigma_3N={\rm i}(MD-DM).$

Based on the above theorem, we merely need to verify the spectral problem. The time evolution part is similar as above theorem. Thus we omit it.
$\square$

\subsection{The uniform transformation through binary DT}
In this subsection, we consider reduction for binary DT. For convenience, we merely consider the one-fold binary DT, since the N-fold binary DT is nothing but
N-times iteration of one-fold DT. Firstly, we consider how to reduce the binary DT into Zakharov-Shabat dressing operator \cite{ZS}.
If the spectral parameters $\lambda_1\neq\bar{\lambda_1}$, we use the relation
\begin{equation*}\begin{split}
                   \left(\frac{\Phi_1^{\dag}\Lambda\Phi_1}{\lambda_1-\bar{\lambda_1}}\right)_x&={\rm i}\Psi_1^{\dag}\sigma_3\Lambda\Phi_1,\\
                    \left(\frac{\Phi_1^{\dag}\Lambda\Phi_1}{\lambda_1-\bar{\lambda_1}}\right)_t&={\rm i}\left[(\lambda_1+\bar{\lambda_1})\Phi_1^{\dag}\Lambda\sigma_3\Phi_1+\Phi_1^{\dag}\sigma_3\Lambda Q\Phi_1\right].
                 \end{split}
\end{equation*}
It follows that
\begin{equation*}
    \Omega(\Phi_1,\Phi_1)=-{\rm i}\left(\frac{\Phi_1^{\dag}\Lambda\Phi_1}{\lambda_1-\bar{\lambda_1}}\right),
\end{equation*}
Then we have
\begin{equation*}
    \Phi[1]=\left(I+\frac{\bar{\lambda_1}-\lambda_1}{\lambda-\bar{\lambda_1}}\frac{\Phi_1\Phi_1^{\dag}\Lambda}{\Phi_1^{\dag}\Lambda\Phi_1}\right)\Phi,\quad Q[1]=Q+(\lambda_1-\bar{\lambda_1})\left[\sigma_3,\frac{\Phi_1\Phi_1^{\dag}\Lambda}{\Phi_1^{\dag}\Lambda\Phi_1}\right].
\end{equation*}
By this transformation, we can obtain bright soliton, breather and rogue wave solution. The high order DT for this kinds of DT was obtain in reference \cite{GLL} by limit technique.

If the spectral parameters $\lambda_1=\bar{\lambda_1}.$ In this case, we use the limit technique to deal with this problem. Suppose $\Psi_1$ and $\Phi_1$ are two different
solution for Lax pair at $\lambda=\lambda_1$ such that $\Phi_1^{\dag}\Lambda\Psi_1=C_1\neq0$ and $\Phi_1^{\dag}\Lambda\Phi_1=0$, set
$\Phi_2=\Phi_1+\frac{\beta(\lambda_2-\lambda_1)}{C_1}\Psi_1$ ,  then we can obtain
\begin{equation*}\begin{split}
                  \lim_{\lambda_2\rightarrow\lambda_1} \left(\frac{\Phi_1^{\dag}\Lambda\Phi_2}{\lambda_2-\lambda_1}\right)_x&={\rm i}\Phi_1^{\dag}\sigma_3\Lambda\Phi_1,\\
                   \lim_{\lambda_2\rightarrow\lambda_1} \left(\frac{\Phi_1^{\dag}\Lambda\Phi_2}{\lambda_2-\lambda_1}\right)_t&={\rm i}\left[2\lambda_1\Phi_1^{\dag}\Lambda\sigma_3\Phi_1+\Phi_1^{\dag}\sigma_3\Lambda Q\Phi_1\right].
                 \end{split}
\end{equation*}
It follows that
\begin{equation*}
    \Omega(\Phi_1,\Phi_1)=-{\rm i}\lim_{\lambda_2\rightarrow\lambda_1}\left(\frac{\Phi_1^{\dag}\Lambda\Phi_2}{\lambda_2-\lambda_1}\right),
\end{equation*}
Then we have
\begin{equation}\label{dark-dt}
\begin{split}
 \Phi[1]&=\lim_{\lambda_2\rightarrow\lambda_1}\left(I+\frac{\lambda_1-\lambda_2}{\lambda-\lambda_1}\frac{\Phi_1\Phi_1^{\dag}\Lambda}{
    \Phi_1^{\dag}\Lambda\Phi_2}\right)\Phi,  \\
   \quad Q[1]&=Q+\lim_{\lambda_2\rightarrow\lambda_1}\left[\sigma_3,\frac{(\lambda_2-\lambda_1)\Phi_1\Phi_1^{\dag}\Lambda}{
    \Phi_1^{\dag}\Lambda\Phi_2}\right].
\end{split}
\end{equation}
To keep the non-singularity of above transformation, we have $$\mathrm{sign}(\beta)=\mathrm{sign}\left(\lim_{\lambda_2\rightarrow\lambda_1}\left(\frac{\Phi_1^{\dag}\Lambda\Phi_1(\lambda_2)}{\lambda_2-\lambda_1}\right)\right).$$
In the following section, we would like to how to use the above transformation to derive the dark soliton and N-dark soliton for N-component NLS equations.

\section{Dark soliton and N-dark soliton} \label{sec3}
In this section, we consider the application for binary Darboux transformation. A direct application is using the Darboux transformation to derive some
special solutions. By the binary Darboux transformation, we can obtain the dark soliton for dCNLS.
\subsection{Single Dark soliton for N-component NLS equation}
To obtain the dark soliton, we use the seed solutions
\begin{equation}\label{seed}
    q_i=c_i\,\exp{\left[{{\rm i} \left( {a_i}\,x- \left( \frac{{{
a_i}}^{2}}{2}+\sum_{l=1}^N\sigma_l{{ c_l}}^{2}\right) t \right) }\right]}.
\end{equation}
In the first place, we need to solve the Lax pair equation
\eqref{L-P} with above seed solutions. In order to solve the Lax
pair equation, we use the gauge transformation
$$D=\mathrm{diag}\left(1,\,\exp{\left[{{\rm i} \left( {a_1}\,x- \left( \frac{{{
a_1}}^{2}}{2}+\sum_{l=1}^N\sigma_l{{ c_l}}^{2}\right) t \right)
}\right]},\cdots\,,\exp{\left[{{\rm i} \left( {a_N}\,x- \left(
\frac{{{ a_N}}^{2}}{2}+\sum_{l=1}^N\sigma_l{{ c_l}}^{2}\right) t \right)
}\right]}\right)$$ converts the variable coefficient differential
equation into constant coefficient equation. Then we can obtain
\begin{equation}\label{L-P-0}
    \begin{split}
      \Phi_{0,x}&={\rm i}U_0\Phi_0, \quad D\Phi=\Phi_0, \\
      \Phi_{0,t}&={\rm i}\left(\frac{1}{2}U_0^2+\lambda {U_0}+\sum_{l=1}^N\sigma_l{{c_l}}^{2}-\frac{1}{2}\,{\lambda}^{2}\right)\Phi_0,
    \end{split}
\end{equation}
where
\begin{equation}\label{U0}
    U_0=\begin{bmatrix}
          \lambda & -CS \\
          C^T &-\lambda I-D  \\
        \end{bmatrix},\quad D=\mathrm{diag}(a_1,\cdots,a_N),\quad
        C=\left(c_1,\,\cdots,c_N\right)
\end{equation}
In the following, we consider the property of the matrix $U_0$. Firstly we can obtain the characteristic equation of matrix $U_0$:
\begin{equation}\label{cubic}
   \det(\mu-U_0)=0.
\end{equation}
Then we have the vector solution for \eqref{L-P-0}:
\begin{equation*}
    \Phi_{0}=\begin{bmatrix}
                 1 \\
                 (\lambda+a_1+\mu)^{-1} \\
                 \vdots \\
                 (\lambda+a_N+\mu)^{-1} \\
               \end{bmatrix}\exp\left[{\rm i}\left( \mu\,x+ \left( \lambda\mu-\frac{1}{2}\,{\lambda}^{2}+\frac{1}{2}\,{\mu}^{2}+\sum_{l=1}^N\sigma_l{c_l}^2\right)t\right)\right].
\end{equation*}

To obtain the dark soliton, we need to choose the complex conjugation root of the characteristic equation \eqref{cubic}.
If $\mu$ and $\bar{\mu}$ are the roots of characteristic equation \eqref{cubic}, then we have
\begin{equation*}
    \mu_j-\lambda_j+\sum_{l=1}^N\frac{\sigma_l{c_l}^2}{\lambda_j+\mu_j+a_l}=0,
\end{equation*}
and
\begin{equation*}
    \bar{\mu_i}-\lambda_i+\sum_{l=1}^N\frac{\sigma_l{c_l}^2}{\lambda_i+\bar{\mu_i}+a_l}=0,
\end{equation*}
where $\mu_j$ and $\bar{\mu_i}$ are the roots of characteristic equation \eqref{cubic} with $\lambda=\lambda_j$ and $\lambda=\lambda_i$ respectively.
It follows that
\begin{equation*}
    \mu_j-\lambda_j-(\bar{\mu_i}-\lambda_i)+\sum_{l=1}^N\frac{\sigma_l{c_l}^2[(\lambda_j+\mu_j)
    -(\lambda_i+\bar{\mu_i})]}{(\lambda_j+\mu_j+a_l)(\lambda_i+\bar{\mu_i}+a_l)}=0.
\end{equation*}
Then we can obtain that
\begin{equation}\label{identi}
    \frac{ \mu_j-\lambda_j-(\bar{\mu_i}-\lambda_i)}{[(\lambda_j+\mu_j)-(\lambda_i+\bar{\mu_i})]}
    +\sum_{l=1}^N\frac{\sigma_l{c_l}^2[(\lambda_j+\mu_j)
    -(\lambda_i+\bar{\mu_i})]}{(\lambda_j+\mu_j+a_l)(\lambda_i+\bar{\mu_i}+a_l)}=0.
\end{equation}
Thus the formula
\begin{equation*}
    \frac{\Phi_j^{\dag}\sigma_3\Phi_i}{\lambda_i-\lambda_j}=\frac{2{\rm e}^{{\rm i}(X_i-\bar{X_j})}}{\lambda_i-\lambda_j+\mu_i-\bar{\mu_j}},
\end{equation*}
where
\begin{equation*}
    \begin{split}
   X_i&=\left[ \mu_i\,x+ \left( \lambda_i\mu_i+\frac{1}{2}\,{\mu_i}^{2}-\frac{1}{2}\,{\lambda_i}^{2}\right)t\right],  \\
   \bar{X_j}&=\left[ \bar{\mu_j}\,x+ \left( \lambda_j\bar{\mu_j}+\frac{1}{2}\,{\bar{\mu_j}}^{2}-\frac{1}{2}\,{\lambda_j}^{2}\right)t\right].
    \end{split}
\end{equation*}
With this formula, we can readily take limit $\lambda_j\rightarrow
\lambda_i$
\begin{equation*}
    \lim_{\lambda_j\rightarrow \lambda_i}\frac{\Phi_j^{\dag}\Lambda\Phi_i}{\lambda_i-\lambda_j}=\frac{2{\rm e}^{{\rm i}(X_i-\bar{X_i})}}{\mu_i-\bar{\mu_i}}.
\end{equation*}
Then we come back to the DT \eqref{dark-dt}. Through above explicit expression and set $\beta=\frac{2{\rm e}^{\alpha}}{\mu_1-\bar{\mu_1}}$, we have
\begin{equation*}
    \lim_{\lambda_2\rightarrow\lambda_1}\frac{\Phi_1^{\dag}\Lambda\Phi_2}{\lambda_2-\lambda_1}=\frac{2({\rm e}^{{\rm i}(X_1-\bar{X_1})}+{\rm e}^{\alpha})}{\mu_1-\bar{\mu_1}}.
\end{equation*}
Then the DT \eqref{dark-dt} can be constructed explicitly as
\begin{equation*}
    T=I-\frac{(\mu_1-\bar{\mu_1})\Phi_1\Phi_1^{\dag}\Lambda}{2(\lambda-\lambda_1)({\rm e}^{{\rm i}(X_1-\bar{X_1})}+{\rm e}^{\alpha})}.
\end{equation*}
It follows that
single Dark soliton for N-component NLS equation is
\begin{equation*}
      q_i[1]=c_i\left[1-\frac{B_i}{2}+\frac{B_i}{2}\tanh(Y_1)\right]\,\exp\left[{{\rm i} \left( {a_i}\,x- \left( \frac{{{
a_i}}^{2}}{2}+{\sum_{l=1}^N\sigma_l{ c_l}^{2}} \right) t \right) }\right],
\end{equation*}
where
\begin{equation*}
    \begin{split}
      B_i&=\frac{\mu_1-\bar{\mu_1}}{\lambda_1+a_i+\mu_1},\quad i=1,2,\cdots,N,
 \\
      Y_1&=-\mathrm{Im}(\mu_1)\left[x+(\lambda+\mathrm{Re}(\mu_1))t+\alpha\right].
    \end{split}
\end{equation*}
Without loss of generality, we suppose $\mathrm{Im}(\mu_1)>0$. When
$x\rightarrow -\infty$, we have
\begin{equation*}
      q_i[1]\rightarrow c_i\,{\exp\left[{{\rm i} \left( {a_i}\,x- \left( \frac{{{
a_i}}^{2}}{2}+\sum_{l=1}^N\sigma_l{{ c_l}}^{2}\right) t \right) }\right]}.
\end{equation*}
When $x\rightarrow +\infty$, we have
\begin{equation*}
       q_i[1]\rightarrow c_i\,{\exp\left[{{\rm i} \left( {a_i}\,x- \left( \frac{{{
a_i}}^{2}}{2}+\sum_{l=1}^N\sigma_l{{ c_l}}^{2}\right) t+\theta_i \right)
}\right]},\quad \ln
\frac{\lambda+a_i+\bar{\mu_1}}{\lambda+a_2+\mu_1}={\rm i}\theta_i.
\end{equation*}

The center of dark-dark soliton $q_i[1]$ is along the line
$x+[\lambda+\mathrm{Re}(\mu_1)]t+\alpha=0.$ The velocity of
dark-dark soliton $q_i[1]$ is $v=-(\lambda_1+\mathrm{Re}(\mu_1)).$ The
depth of cavity of $|q_i[1]|^2$ is
$$\frac{c_i^2\mathrm{Im}^2(\mu_1)}{(\lambda+a_i+\mathrm{Re}(\mu_1))^2+\mathrm{Im}^2(\mu_1)}.$$

In the following, we introduce another way to determine the
background whether or not exist the dark-dark bound states \cite{OWY}. Through the relation \eqref{identi}, we have
\begin{equation}\label{velocity}
    \sum_{l=1}^N\frac{\sigma_l{c_l}^2}{|\lambda_1+\mu_1+a_l|^2}=1.
\end{equation}
Indeed, through the expression for single soliton, to obtain the need to know the parameter $\lambda_1+\mu_1$. And the velocity of soliton is controlled by
$-\mathrm{Re}(\lambda_1+\mu_1).$ Thus if we need to find the soliton with velocity equals to zero, we merely to solve the following equation.
\begin{equation}\label{zero}
    \sum_{l=1}^N\frac{\sigma_l{c_l}^2}{a_l^2+\mathrm{Im}^2(\mu_1)}=1.
\end{equation}
If $\sigma_l=1$ for all $l$, then the function
$F(\mathrm{Im}(\mu_1))=\sum_{l=1}^N\frac{\sigma_l{c_l}^2}{a_l^2+\mathrm{Im}^2(\mu_1)}$
is an increasing function in positive half axis. Then the equation
\eqref{zero} merely has a positive solution. Thus in defocusing case,
there no exists bound state. So the multi-bound state merely exists in the mixed case.

\begin{figure}[htb]
\centering
\includegraphics[height=50mm,width=80mm]{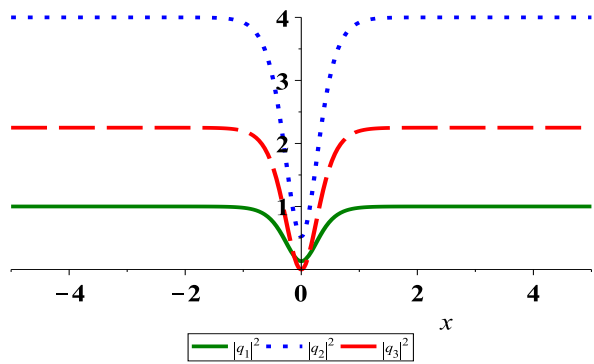}
\caption{(color online): $t=0$: Solid green line $|q_1|^2$,  Dot blue line
$|q_2|^2$, Dash red line $|q_3|^2$.}\label{fig1}
\end{figure}

In what following, we illustrate some exact examples to the single dark soliton.  Since the velocity of soliton possesses the exact physical meaning, we look for the soliton by the velocity $v=-\mathrm{Re}(\lambda_1+\mu_1)$.  Then we can solve the following equation about $\mathrm{Im}(\mu_1)$
 \begin{equation*}
    \sum_{l=1}^N\frac{\sigma_l{c_l}^2}{(a_l-v)^2+\mathrm{Im}^2(\mu_1)}=1.
 \end{equation*}
Substituting $\mu_1=-(v+\lambda_1)+{\rm i}\mathrm{Im}^2(\mu_1)$ into characteristic equation \eqref{coeff1}, we can obtain an algebraic equation about $\lambda_1$. Solving the algebraic equation, we can obtain all of parameters about single dark soliton.
 For instance, we consider the three-component NLS with the defocusing case (i.e. $N=3$, $\sigma_1=\sigma_2=\sigma_3=1$). If we need to find the soliton with
 velocity $v=0$, then we choose the parameters:
\begin{equation*}
    \begin{split}
       a_1&=1,\,a_2=-1,\,a_3=0,\,c_1=1,\,c_2=2,\,c_3=\frac{3}{2},\\
       \lambda_1&=-\frac{12}{33+\sqrt{769}},\, \mu_1=\frac{99-3\sqrt{769}}{80}+\frac{{\rm
i}}{4}\sqrt{50+2\sqrt{769}}.
    \end{split}
\end{equation*}
We can plot the picture of single dark soliton by Maple (Fig. \ref{fig1}). Since the soliton are stationary, we merely plot the picture at $t=0$.

\subsection{N-dark soliton for N-component NLS equation}
In order to give the N-dark soliton, we first adapt the binary DT with the limit technique.
The N-fold binary DT can be written with the following form
\begin{equation}\label{N-fold}
    \Phi[N]=\Phi-\sum_{i=1}^Ns_i\Omega(\Phi_i,\Phi).
\end{equation}
Thus we can suppose that
\begin{equation}\label{N-fold}
    \Phi[N]=T_N\Phi,\quad T_N=I-\sum_{i=1}^Ns_i\frac{\Phi_i^{\dag}\Lambda}{\lambda-\lambda_i}.
\end{equation}
The explicit expression for Darboux matrix $T_N$ can be determined by the following equations
\begin{equation*}
    \lim_{\lambda\rightarrow\lambda_j}T_N\left(\Phi_j+\frac{\beta_j}{C_j}(\lambda-\lambda_j)\Psi_j\right)=0,
\end{equation*}
where $\beta_j=\frac{2{\rm e}^{\alpha_j}}{\mu_j-\bar{\mu_j}},$ $C_j=\Phi_j^{\dag}\Lambda \Psi_j.$
By linear algebra, we have the following expression for $T_N$:
\begin{equation}\label{TN}
    T_N=I-YM^{-1}(\lambda-D)^{-1}Y^{\dag}\Lambda,\quad Y=\begin{bmatrix}
                                                   \Phi_1\,,& \Phi_2\,, &\cdots\,,  & \Phi_N \\
                                                  \end{bmatrix}
\end{equation}
where
\begin{equation*}
    M=\begin{bmatrix}
        \lim_{\lambda\rightarrow\lambda_1}\frac{\Phi_1^{\dag}\Lambda\Phi}{\lambda-\lambda_1}+\beta_1 & \frac{\Phi_1^{\dag}\Lambda\Phi_2}{\lambda_2-\lambda_1} & \cdots & \frac{\Phi_1^{\dag}\Lambda\Phi_N}{\lambda_N-\lambda_1} \\
        \frac{\Phi_2^{\dag}\Lambda\Phi_1}{\lambda_1-\lambda_2} &  \lim_{\lambda\rightarrow\lambda_2}\frac{\Phi_2^{\dag}\Lambda\Phi}{\lambda-\lambda_2}+\beta_2 &  \cdots & \frac{\Phi_2^{\dag}\Lambda\Phi_N}{\lambda_N-\lambda_2}  \\
       \vdots  & \vdots & \ddots & \vdots \\
       \frac{\Phi_N^{\dag}\Lambda\Phi_1}{\lambda_1-\lambda_N}  & \frac{\Phi_N^{\dag}\Lambda\Phi_2}{\lambda_2-\lambda_N}   & \cdots &  \lim_{\lambda\rightarrow\lambda_N}\frac{\Phi_N^{\dag}\Lambda\Phi}{\lambda-\lambda_N}+\beta_N \\
      \end{bmatrix}.
\end{equation*}

\begin{figure}[htb]
\centering
\subfigure[$|q_1|^2$]{\includegraphics[height=40mm,width=40mm]{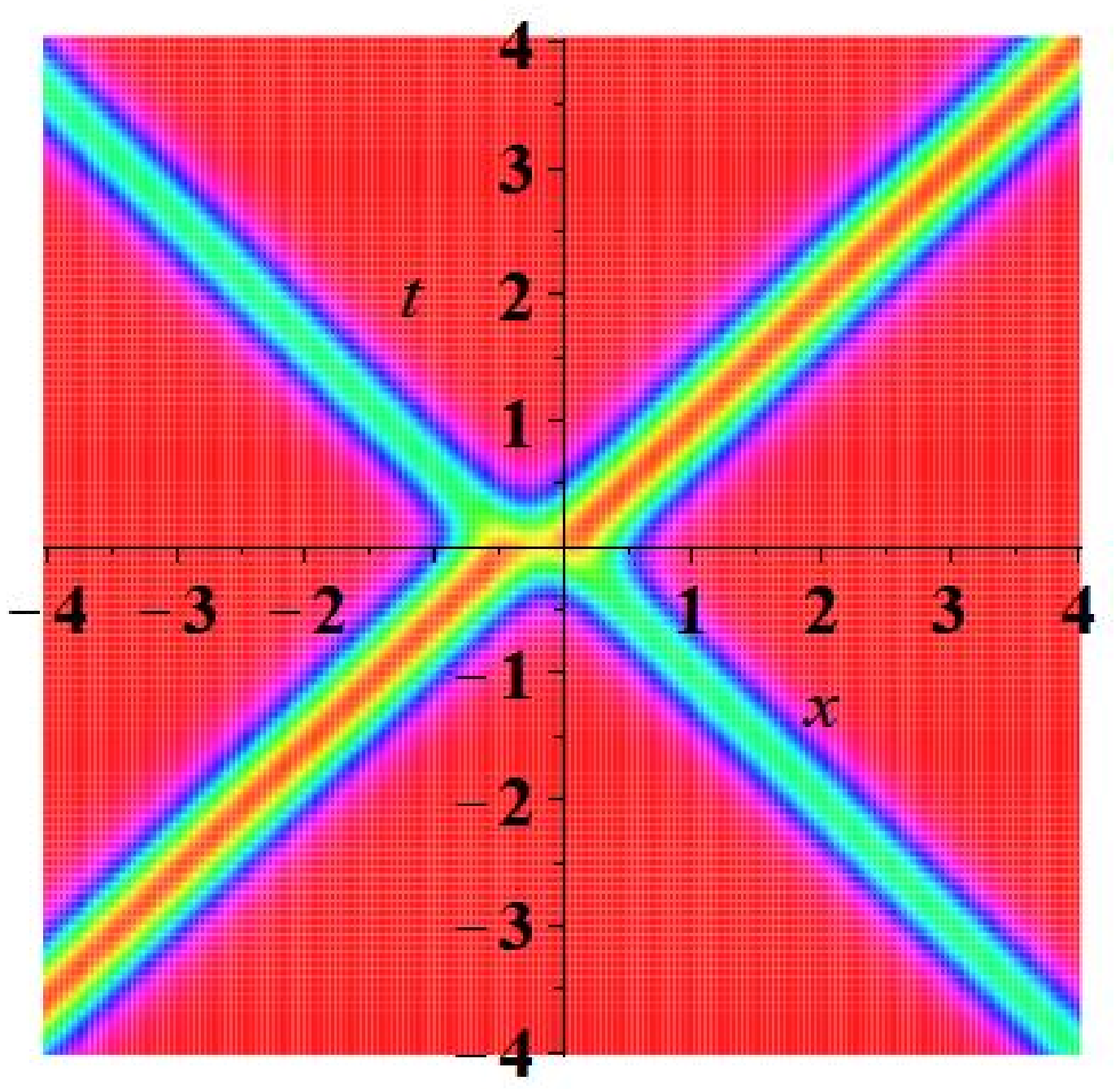}}
\hfil
\subfigure[$|q_2|^2$]{\includegraphics[height=40mm,width=40mm]{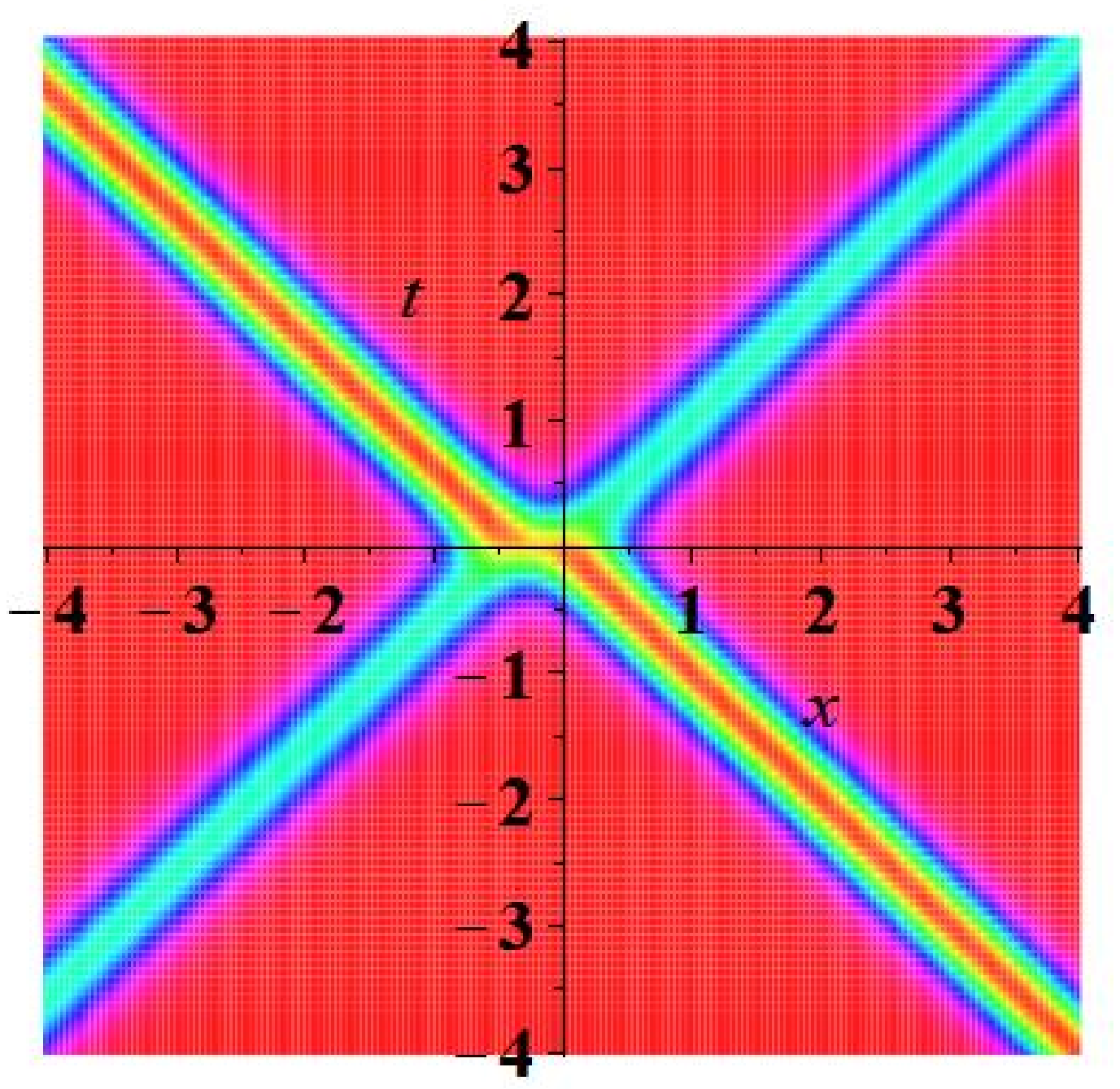}}
\hfil
\subfigure[$|q_3|^2$]{\includegraphics[height=40mm,width=40mm]{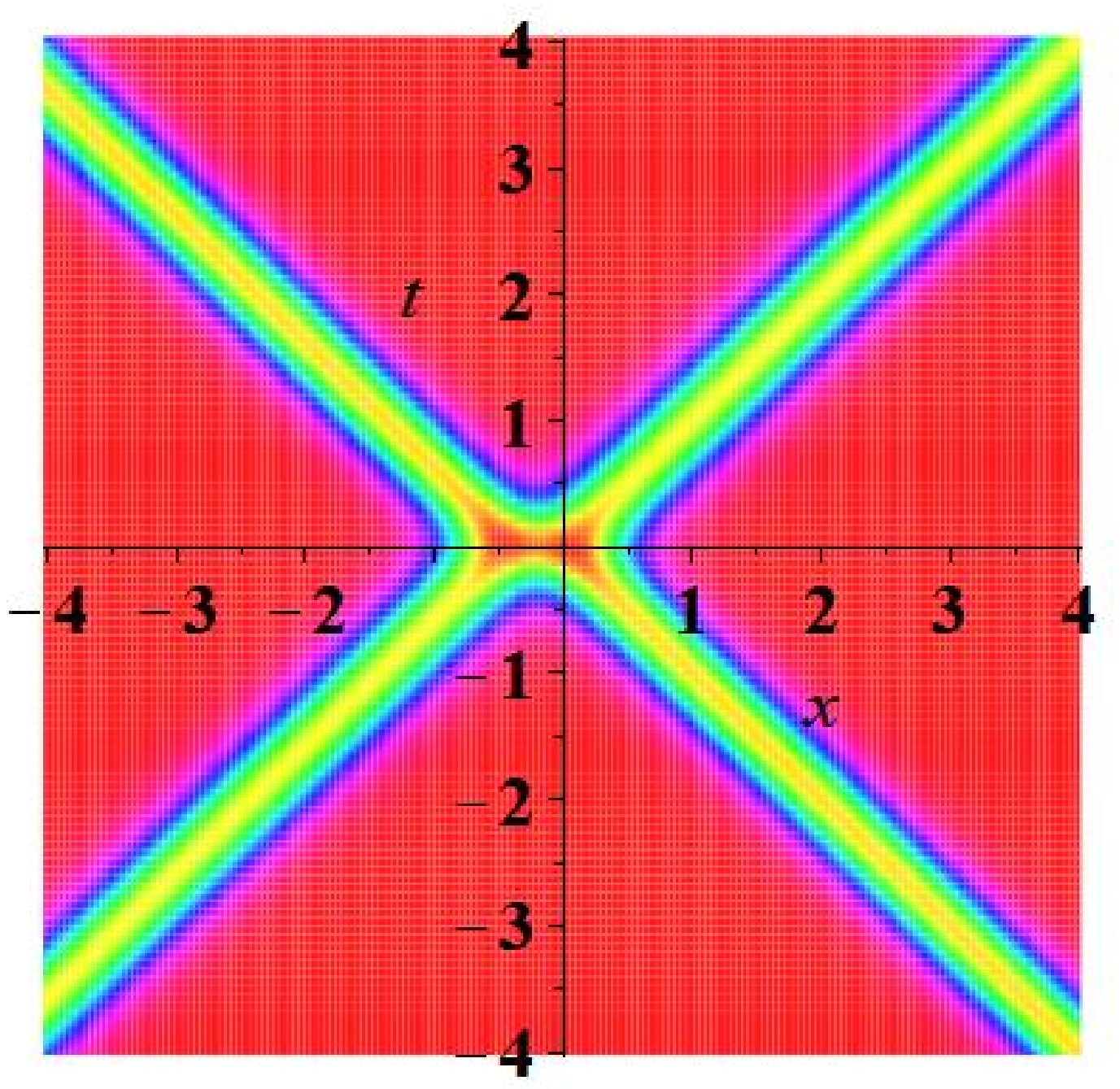}}
\hfil
\subfigure[$t=-2$]{\includegraphics[height=50mm,width=70mm]{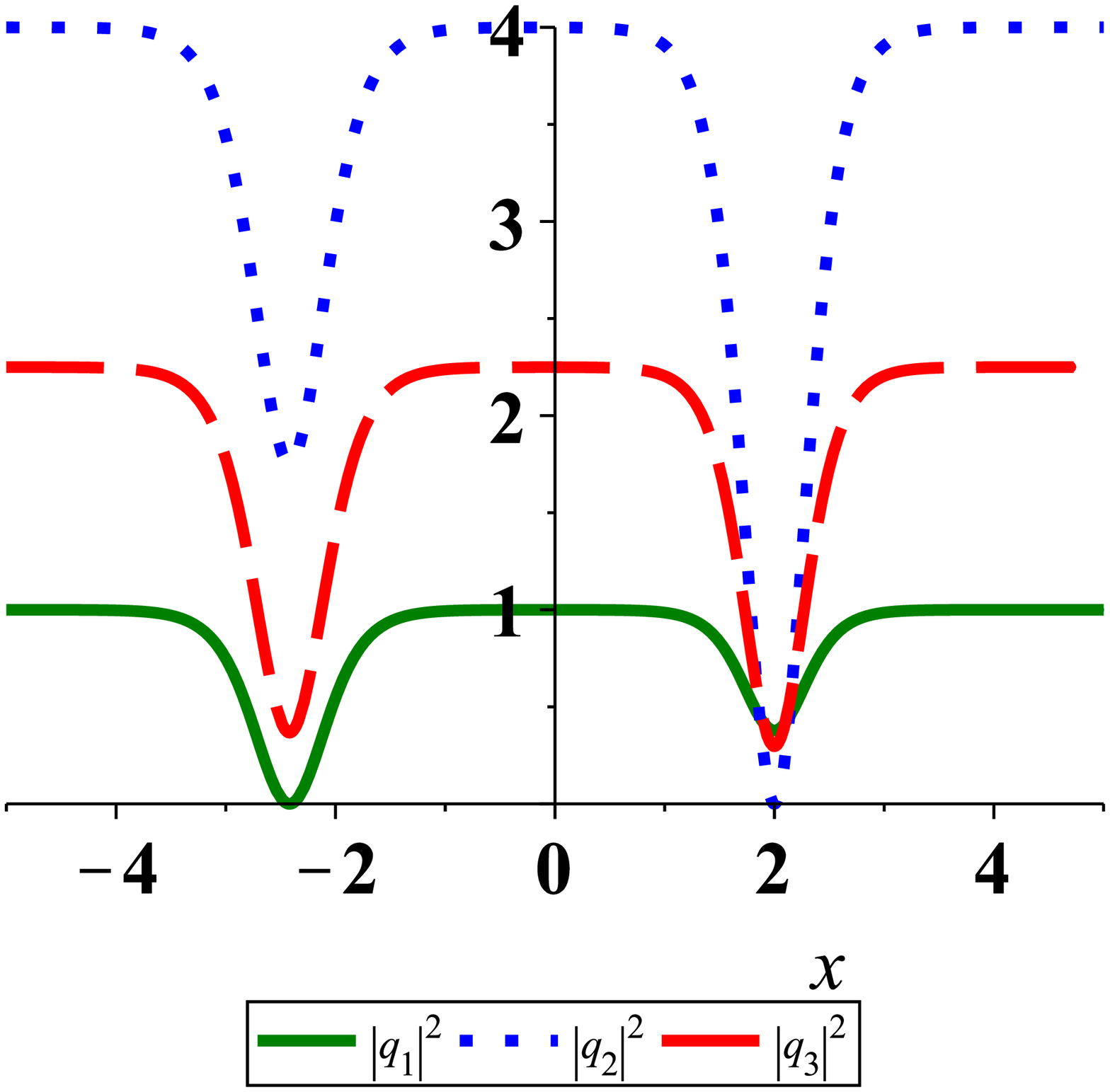}}
\hfil
\subfigure[$t=2$]{\includegraphics[height=50mm,width=70mm]{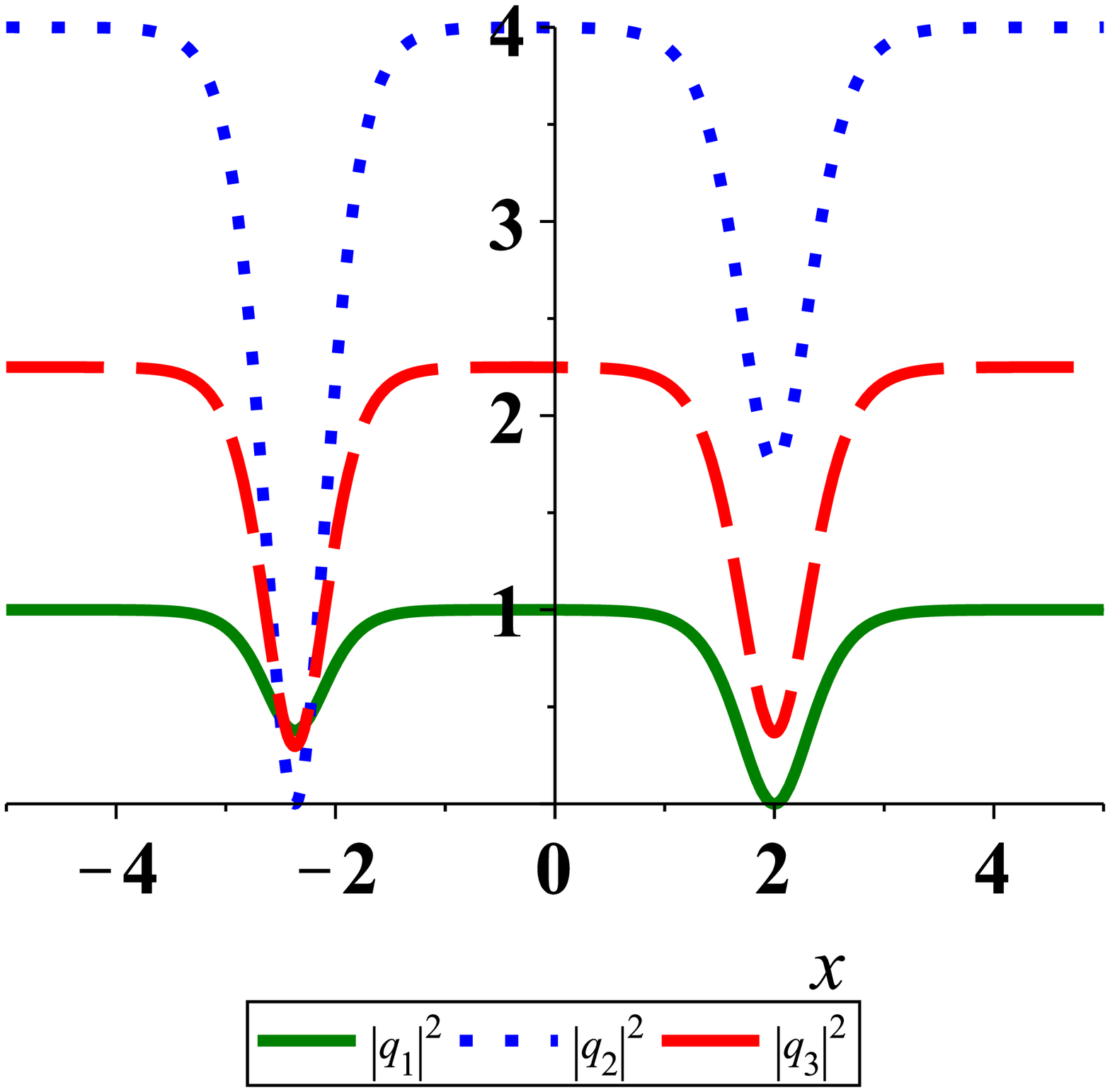}}
\caption{(color online) (a)-(c), Density plot of $|q_1|^2$,
$|q_2|^2$ and $|q_3|^2$ respectively; (d),(e) Solid green line $|q_1|^2$, Dot
blue line $|q_2|^2$, Dash red line $|q_3|^2$. (d), $t=-2$, (e), $t=2$. }\label{fig2}
\end{figure}

By the above subsection, we have
\begin{equation*}
    \frac{\Phi_i^{\dag}\Lambda\Phi_j}{\lambda_j-\lambda_i}=\frac{2{\rm e}^{{\rm
    i}(X_j-\bar{X_i})}}{(\lambda_j+\mu_j)-(\lambda_i+\bar{\mu_i})}.
\end{equation*}
 With this formula, we can readily take limit $\lambda\rightarrow
\lambda_i$
\begin{equation*}
    \lim_{\lambda\rightarrow \lambda_i}\frac{\Phi_i^{\dag}\Lambda\Phi}{\lambda-\lambda_i}=\frac{2{\rm e}^{{\rm i}(X_i-\bar{X_i})}}{\mu_i-\bar{\mu_i}}.
\end{equation*}
Then the N-Dark soliton for equations \eqref{dcnls} can be represented as following:
\begin{equation*}
       q_i[N]=2\frac{\begin{vmatrix}
                       M & X^{\dag} \\
                       Y_i & \frac{c_i}{2} \\
                     \end{vmatrix}
       }{|M|}\exp\left[{{\rm i} \left( {a_i}\,x- \left( \frac{{{
a_i}}^{2}}{2}+\sum_{l=1}^N\sigma_l{c_l}^2\right) t\right) }\right],
\end{equation*}
where
\begin{equation*}
   \begin{split}
      M&=\left(\frac{2\left[{\rm e}^{{\rm i}(X_j-\bar{X_i})}+\delta_{ij}\right]}{(\lambda_j+\mu_j)-(\lambda_i+\bar{\mu_i})}\right)_{1\leq i,j\leq N},\quad \delta_{ij}=\left\{
                    \begin{array}{ll}
                      0, & i\neq j, \\
                      e^{\alpha_i}, & i=j,
                    \end{array}
                  \right.
       \\
     X &=\left[{\rm e}^{{\rm i}X_1},{\rm e}^{{\rm i}X_2},\cdots,{\rm e}^{{\rm i}X_N}\right],\\
     Y_i &=\left[\beta_{1,i}{\rm e}^{{\rm i}X_1},\beta_{2,i}{\rm e}^{{\rm i}X_2},\cdots,\beta_{N,i}{\rm e}^{{\rm i}X_N}\right],
     \quad\beta_{j,i}=\frac{1}{\lambda_j+a_i+\mu_j}.
   \end{split}
\end{equation*}
In what following, we consider some dynamics for two dark soliton
for three-component NLS equation. Firstly, we consider the
defocusing case $\sigma_i=1$, $i=1,2,3$. By the method in the above subsection (we choose the velocity $v_1=1$ and $v_2=-1$), the parameters are choosing as following: \begin{equation*}
\begin{split}
 a_1&=1,\,a_2=-1,\,a_3=0,\,c_1=1,\,c_2=2,\,c_3=\frac{3}{2},\\ \lambda_1&=-1.121588903,\,\lambda_2=0.7430861497,\,\alpha_1=\alpha_2=0,\\
 \mu_1&=0.1215889040+2.265094396{\rm i},\,\mu_2=0.2569138501+2.564117194{\rm
 i}.
\end{split}
\end{equation*}
Then we can obtain figure \ref{fig2}.

\begin{figure}[htb]
\centering
\subfigure[$|q_1|^2$]{\includegraphics[height=40mm,width=40mm]{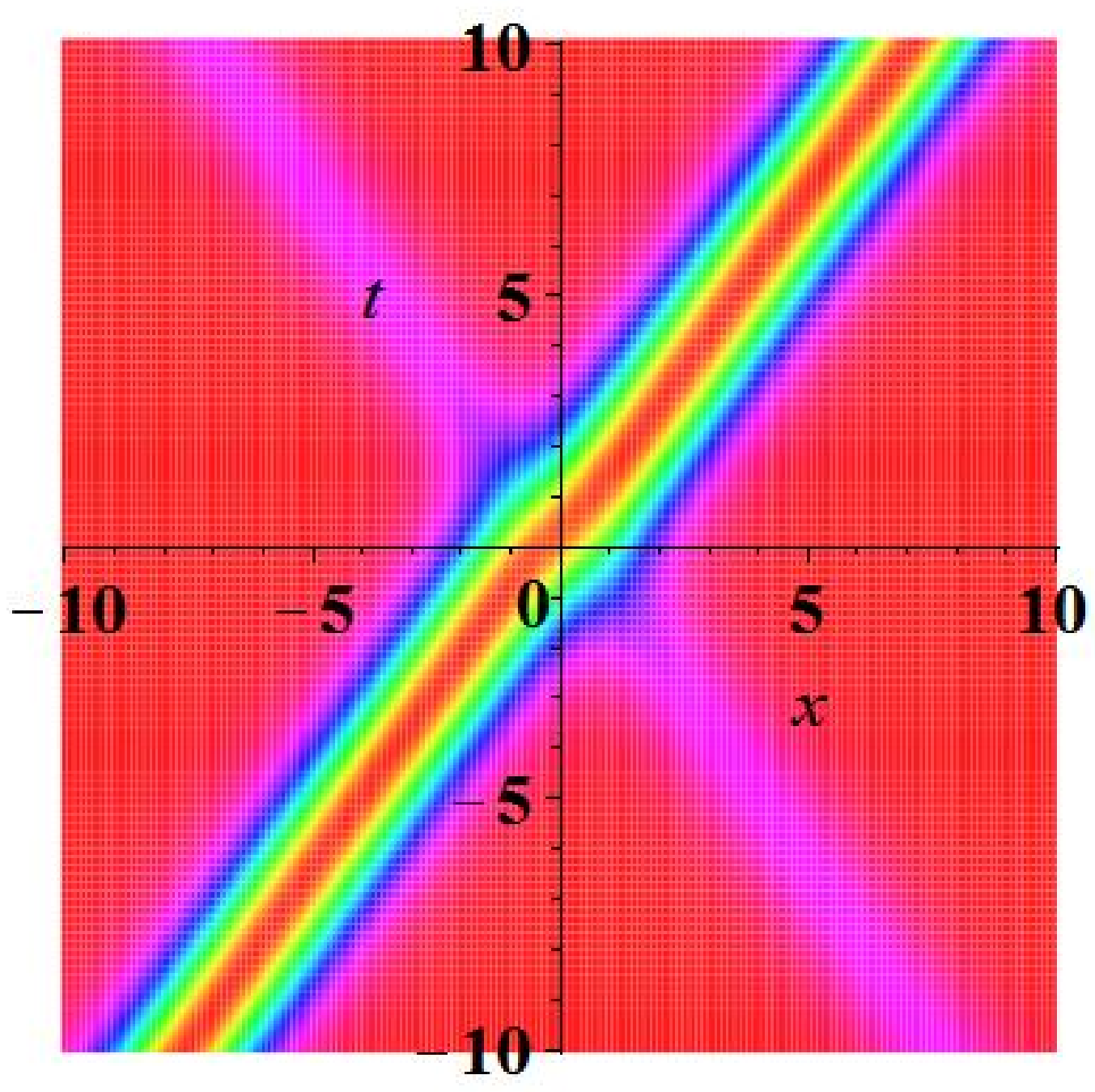}}
\hfil
\subfigure[$|q_2|^2$]{\includegraphics[height=40mm,width=40mm]{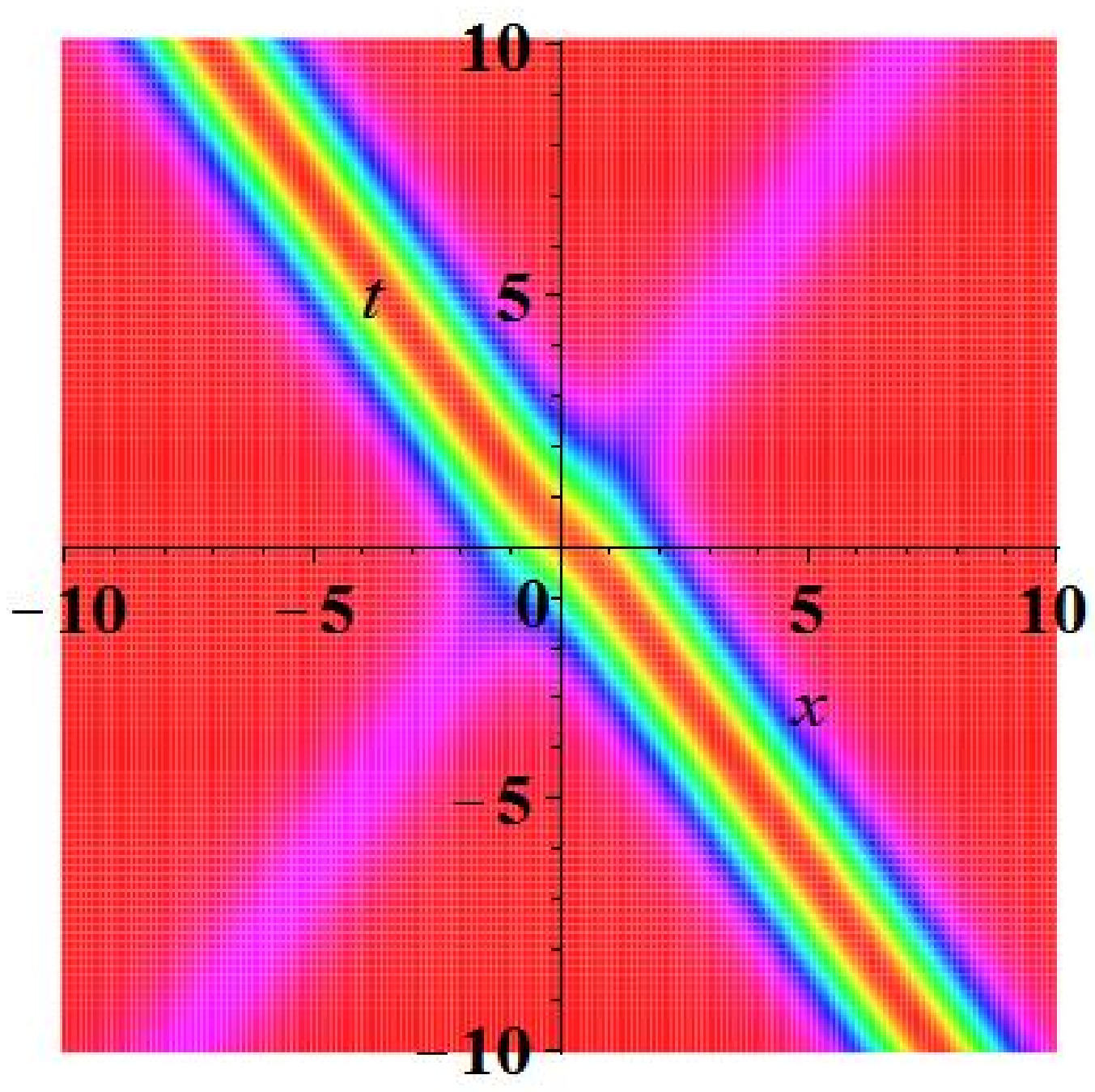}}
\hfil
\subfigure[$|q_3|^2$]{\includegraphics[height=40mm,width=40mm]{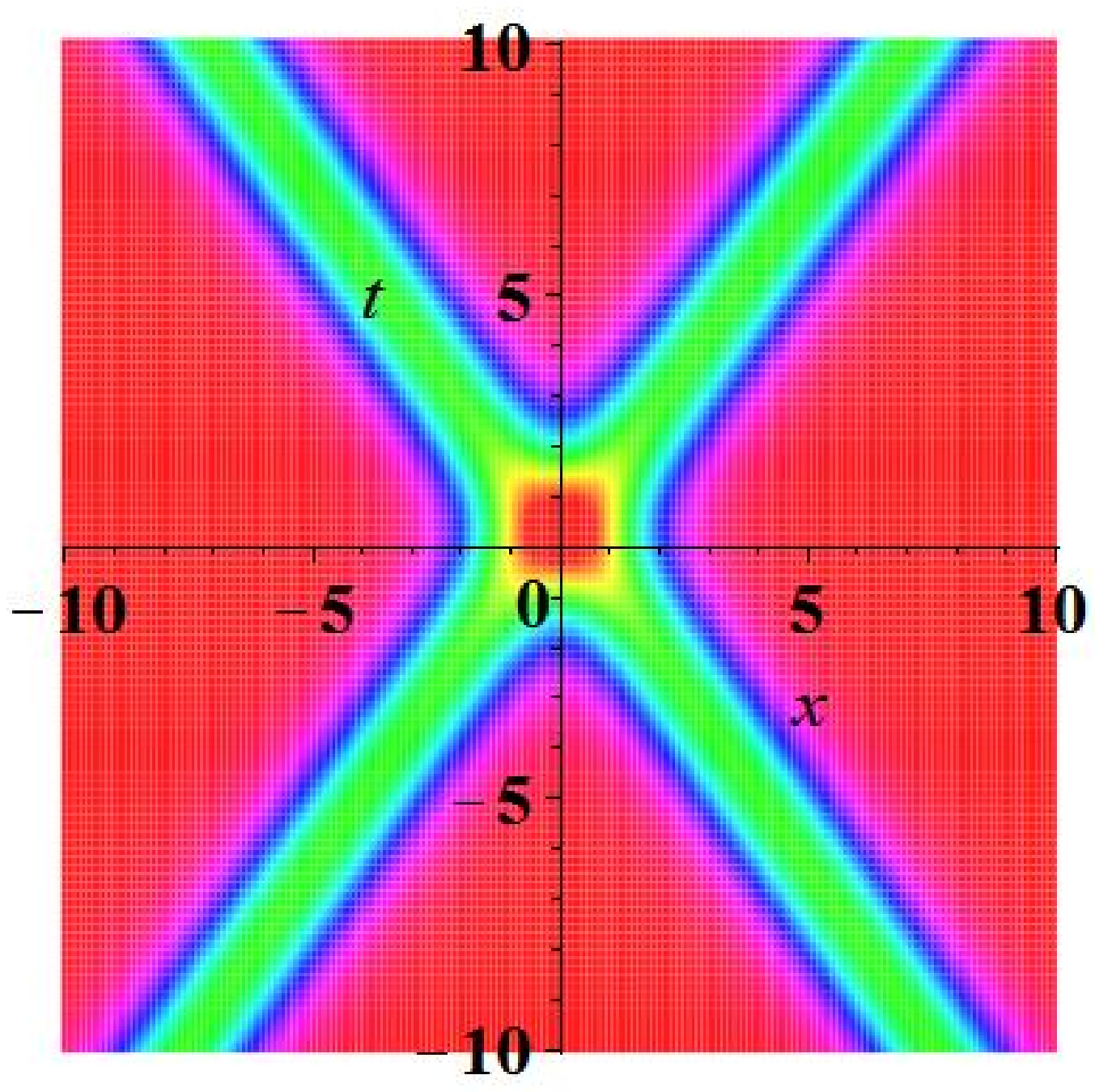}}
\hfil
\subfigure[$t=-5$]{\includegraphics[height=50mm,width=70mm]{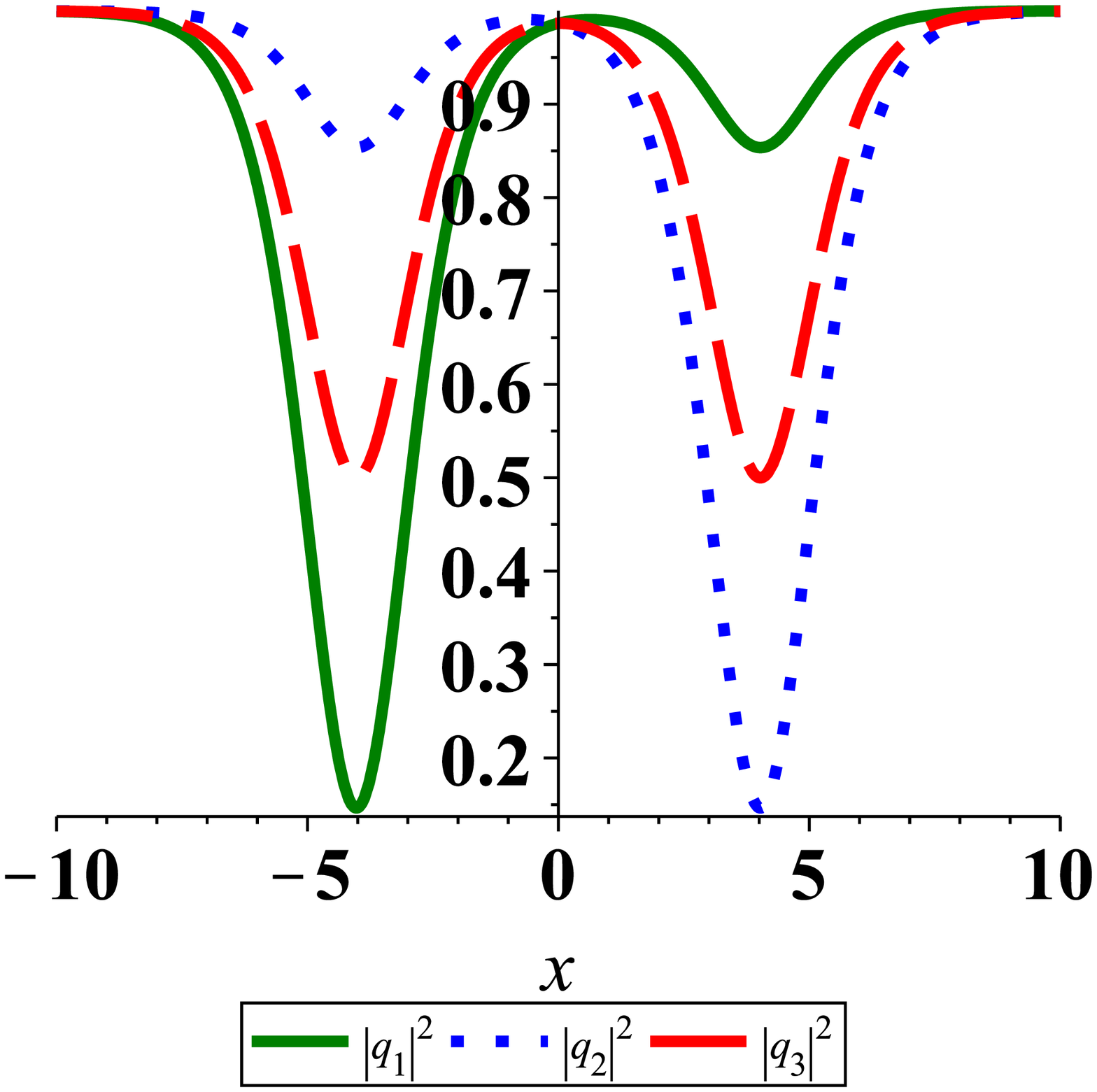}}
\hfil
\subfigure[$t=5$]{\includegraphics[height=50mm,width=70mm]{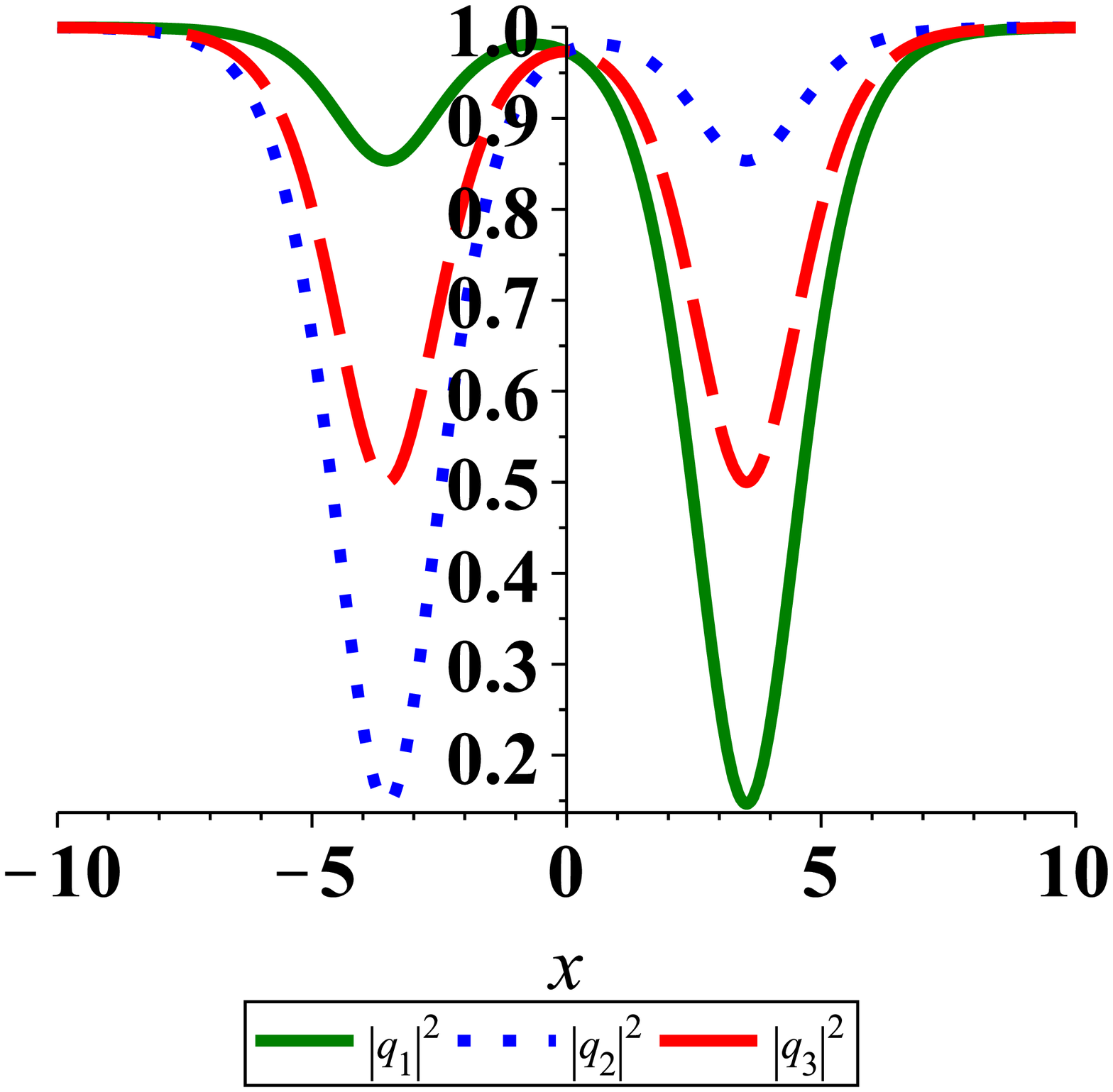}}
\caption{(color online) (a)-(c),Density plot of $|q_1|^2$, $|q_2|^2$
and $|q_3|^2$; (d),(e) solid green line $|q_1|^2$, dot blue line
$|q_2|^2$, dash red line $|q_3|^2$, }\label{fig3}
\end{figure}

\begin{figure}[htb]
\centering
\subfigure[$|q_1|^2$]{\includegraphics[height=40mm,width=40mm]{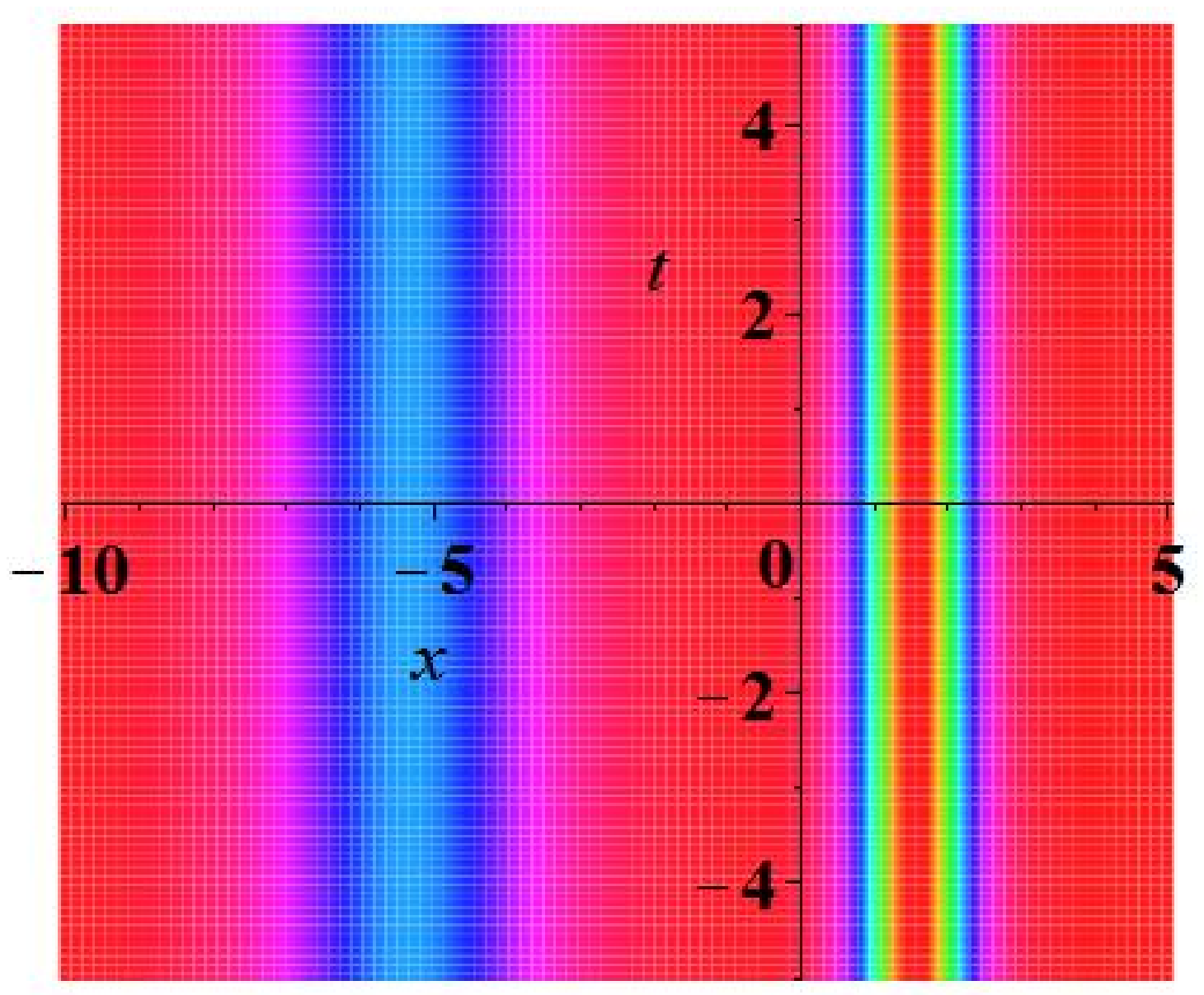}}
\hfil
\subfigure[$|q_2|^2$]{\includegraphics[height=40mm,width=40mm]{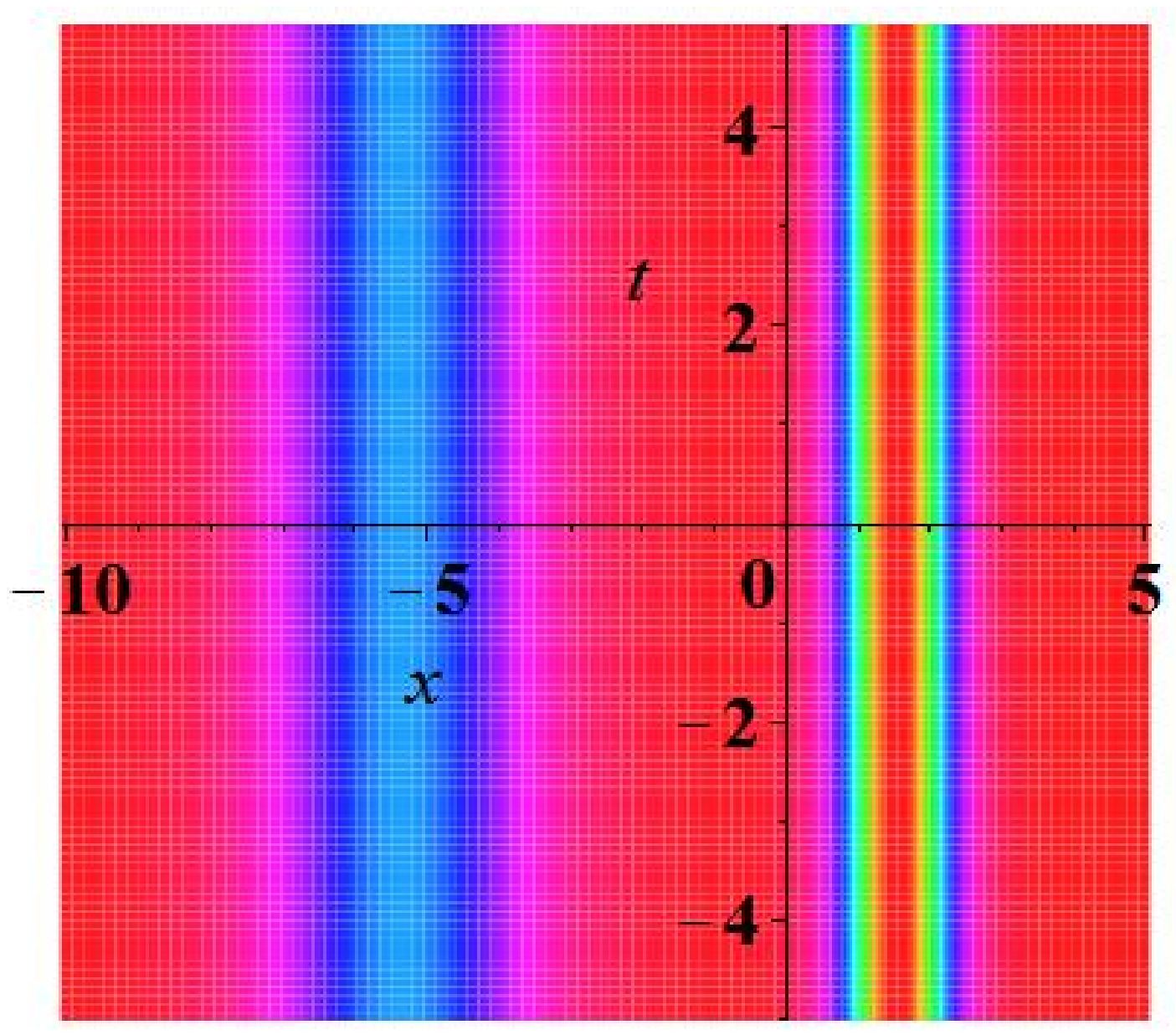}}
\hfil
\subfigure[$|q_3|^2$]{\includegraphics[height=40mm,width=40mm]{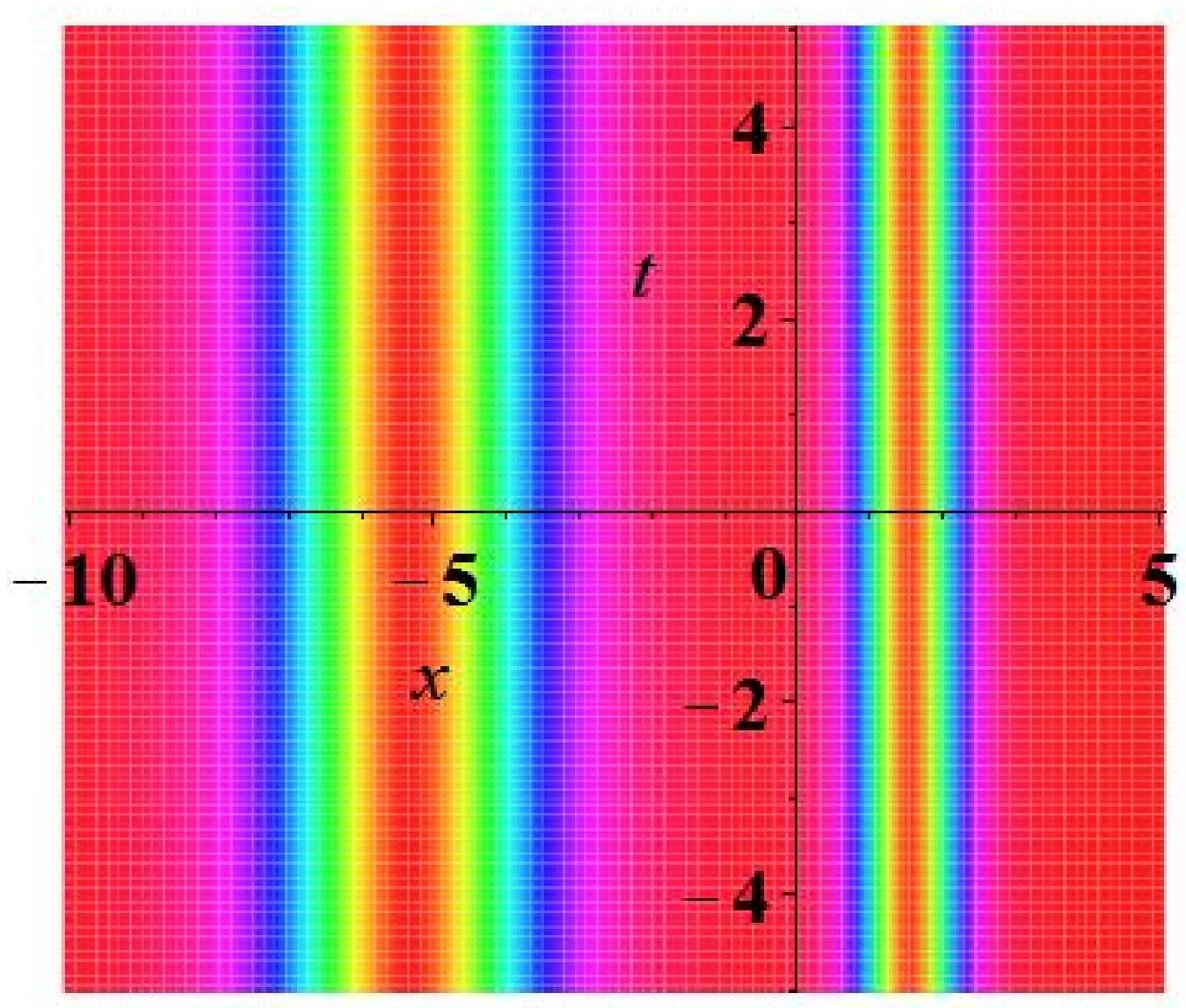}}
\hfil
\subfigure[$t=0$]{\includegraphics[height=50mm,width=80mm]{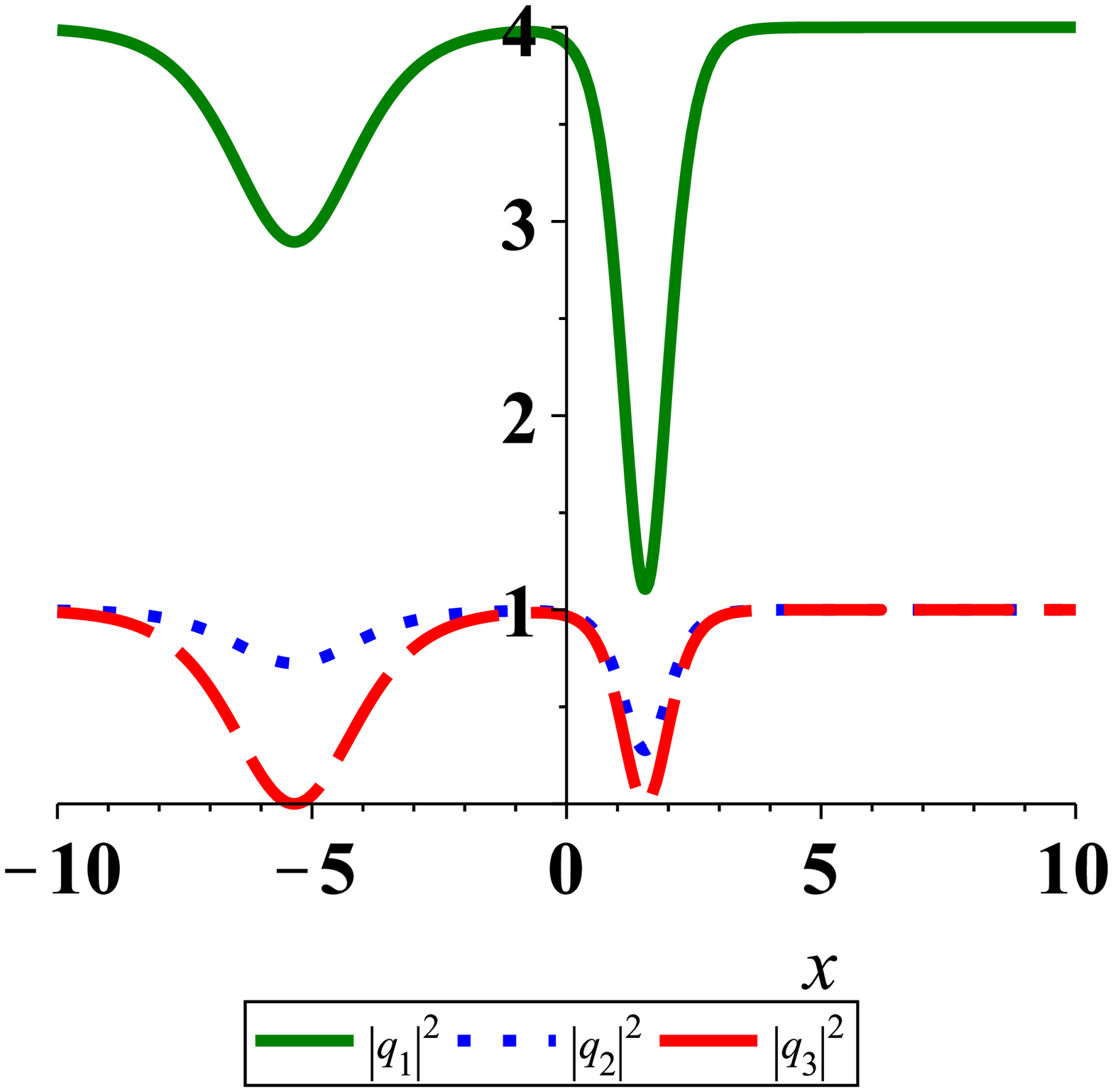}}
\caption{(color online): (a)-(c),Density plot of $|q_1|^2$, $|q_2|^2$
and $|q_3|^2$; (d) Solid green line $|q_1|^2$, Dot blue line
$|q_2|^2$, Dash red line $|q_3|^2$, $t=0$}\label{fig4}
\end{figure}

Secondly, we consider the mixed focusing and defocusing case
$\sigma_i=1$, $i=1,2$ and $\sigma_3=-1$. In the first place, we consider the two-pole two-dark soliton. Since the characteristic equation \eqref{cubic} for three-component NLS equation is a quartic equation, there maybe exists two pairs of complex conjugation roots. This kinds of soliton
can not exist in the scalar or two-component NLS system, since the characteristic equation is not allowed to exist two pairs of conjugation complex roots.
For instance, we choose the parameters as following
\begin{equation*}
\begin{split}
  a_1&=1,\,a_2=-1,\,a_3=0,\,c_1=c_2=c_3=1,  \\
  \lambda_1&=\lambda_2=0,\,\mu_1=\frac{\sqrt{2}}{2}(1+{\rm
    i}),\,\mu_2=-\frac{\sqrt{2}}{2}(1+{\rm
    i}),\,\alpha_1=\alpha_2=0.
\end{split}
\end{equation*}
The figures are giving in Fig \ref{fig3}. It is seen that there is not evident different dynamics behavior with ordinary two-dark soliton.
Then, we consider the two bound state of three-component NLS equations. The parameters are choosing by the method of above subsection. Firstly, we choose the background parameters $a_i$, $c_i$ and $v=0$, we can obtain two different $\mathrm{Im}(\mu_1),\mathrm{Im}(\mu_2)$. And then substitute into \eqref{cubic}, we can obtain the two different spectral parameters $\lambda_1,\lambda_2$. Since the parameter $\alpha_1,\alpha_2$ depend the initial position of soliton, we choose it for different value to distinguish two soliton. For instance, we choose the parameters as following
\begin{equation*}
\begin{split}
  a_1&=1,\,a_2=-1,\,a_3=0,\,c_1=2,\,c_2=c_3=1,\\
  \lambda_1&=\frac{3}{2}\frac{\sqrt{5}-1}{3\sqrt{5}-5},\,\lambda_2=
    \frac{3}{2}\frac{\sqrt{5}+1}{3\sqrt{5}-5},\,\alpha_1=5,\,\alpha_2=-5\\ \mu_1&=-\frac{15+3\sqrt{5}}{20}+\frac{{\rm i}}{2}(\sqrt{5}-1),\,\mu_2=
    -\frac{15-3\sqrt{5}}{20}+\frac{{\rm
    i}}{2}(\sqrt{5}+1),
\end{split}
\end{equation*}
The figures are giving in Fig. \ref{fig4}.

\subsection{Conclusion}
In this paper, we obtain the uniform transformation for
multi-component NLS equation. To the best of our knowledge, the mean
of this transformation is two-fold. The Darboux transformation is
related with inverse scattering transformation which is a method to
solve the initial value problem of integrable PDE. The inverse scattering method of
coupled NLS equation is an open problem in soliton theory. In 2006,
Abolowitz et.al solved this problem with the special background \cite{ABP}. The DT provides a way to solve this problem
 at least for the discrete spectrum.

Besides there is another open question in the famous book of
Faddeev and Takhtajan (P. 145 the end of second paragraph). They authors deem that
the solution of Riemann problem with zeros cannot be expressed as a product of Blaschke-Potapov factors and
a solution of the regular Riemann problem with same continuous spectrum data. Indeed, by the above binary DT, we can delete or add the discrete spectrum for the spectral problem:
\begin{equation*}
   L\Phi=\lambda\Phi,\quad L= {\rm i}\sigma_3\partial_x-\sigma_3 Q,\quad Q=\begin{bmatrix}
                                                                             0 & \bar{q} \\
                                                                             q & 0 \\
                                                                           \end{bmatrix}.
\end{equation*}
By the binary DT, we can readily construct the $L^2(\mathbb{R})$ eigenfunction
\begin{equation*}
    \Phi[1](\lambda_1)=\frac{\beta\Phi_1}{\lim_{\lambda\rightarrow \lambda_1}\frac{\Phi_1^{\dag}\sigma_3\Phi}{\lambda-\lambda_1}+\beta},\quad Q[1]=Q+\lim_{\lambda\rightarrow\lambda_1}\left[\sigma_3,\frac{(\lambda-\lambda_1)\Phi_1\Phi_1^{\dag}\sigma_3}{
    \Phi_1^{\dag}\sigma_3\Phi+(\lambda-\lambda_1)\beta}\right],
\end{equation*}
where $\mathrm{sign}(\lim_{\lambda\rightarrow \lambda_1}\frac{\Phi_1^{\dag}\sigma_3\Phi}{\lambda-\lambda_1})=\mathrm{sign}(\beta).$ It is readily to verify that
$\Phi[1](\lambda_1)\in L^2(\mathbb{R}).$
The detail research for this transformation applied to inverse scattering transformation will be proceeded in the future.

The direct and simple application for this transformation is using it to derive the dark and N-dark soliton solution.
Besides, the method in our paper can be readily to generalize
in the other integrable system to look for some hiding solutions. We would like to explore it in the future.

\bigskip
\noindent {\bf Acknowledgments}

This work is supported by National Natural Science Foundation of
China 11271052.

\end{document}